\RequirePackage{fix-cm}
\documentclass[twocolumn,epjc3]{svjour3}  
\smartqed  
\RequirePackage{graphicx}
\RequirePackage[colorlinks,citecolor=blue,urlcolor=blue,linkcolor=blue]{hyperref} 
\usepackage{amsmath}
\usepackage{color}
\usepackage{ifthen}
\usepackage{multirow}
\newcommand{\ctsper}      {cts/(keV$\cdot$kg$\cdot$yr)}

\newcommand{\mus}         {{$\upmu$s}}

\newcommand{\qbb}         {{$Q_{\beta\beta}$}}

\newcommand{\onbb}        {{$0\nu\beta\beta$}}
\newcommand{\nnbb}        {{$2\nu\beta\beta$}}

\newcommand{\etal}        {\textit{et al.}}

\newcommand{\gerda}       {\textsc{Gerda}}

\newcommand{\GERDA}       {\mbox{\textsc{Gerda}}}  
\newcommand{\igex}        {\textsc{Igex}}
\newcommand{\IGEX}        {{\mbox{\textsc{Igex}}}}

\newcommand{\HdM}         {\mbox{\textsc{HdM}}}

\usepackage{upgreek}
\journalname{Eur. Phys. J. C}
\begin{document}

\title{Pulse shape discrimination for \mbox{\sc{Gerda}} Phase~I data}

\titlerunning{The \textsc{Gerda} PSD analysis}   

\author{
M.~Agostini\thanksref{TUM} \and
M.~Allardt\thanksref{DD} \and
E.~Andreotti\thanksref{GEEL,TU} \and
A.M.~Bakalyarov\thanksref{KU} \and
M.~Balata\thanksref{ALNGS} \and
I.~Barabanov\thanksref{INR} \and
M.~Barnab\'e Heider\thanksref{HD,TUM,nowCAN} \and
N.~Barros\thanksref{DD} \and
L.~Baudis\thanksref{UZH} \and
C.~Bauer\thanksref{HD}
% ...
}

\authorrunning{the \textsc{Gerda} collaboration}

\thankstext{nowCAN}{\emph{Present Address:} CEGEP St-Hyacinthe,
 Qu{\'e}bec, Canada}
\thankstext{nowLNGS}{\emph{Present Address:} INFN  LNGS, Assergi, Italy}
\thankstext{alsoFI}{\emph{also at:} Universit{\`a} di Firenze, Italy}
\thankstext{alsoMIPT}{\emph{also at:} Moscow Inst. of Physics and Technology,
  Russia} 
\thankstext{alsoIUN}{\emph{also at:} Int. Univ. for Nature, Society and
    Man ``Dubna'', Russia} 
\thankstext{nowSJU}{\emph{Present Address:} Shanghai Jiaotong University,
  Shanghai, China} 
\thankstext{nowUNC}{\emph{Present Address:} University North Carolina, Chapel
  Hill, USA} 
\thankstext{AQU}{\emph{Present Address:} University of L'Aquila, Dipartimento
        di Fisica, L'Aquila, Italy}
\thankstext{corrauthor}{\emph{Correspondence},
                                email: gerda-eb@mpi-hd.mpg.de}
\institute{
INFN Laboratori Nazionali del Gran Sasso, LNGS, Assergi, Italy\label{ALNGS} \and
Institute of Physics, Jagiellonian University, Cracow, Poland\label{CR} \and
Institut f{\"u}r Kern- und Teilchenphysik, Technische Universit{\"a}t Dresden,
      Dresden, Germany\label{DD} \and
Joint Institute for Nuclear Research, Dubna, Russia\label{JINR} \and
Institute for Reference Materials and Measurements, Geel,
     Belgium\label{GEEL} \and
Max-Planck-Institut f{\"u}r Kernphysik, Heidelberg, Germany\label{HD} \and
Dipartimento di Fisica, Universit{\`a} Milano Bicocca,
     Milano, Italy\label{MIBF} \and
INFN Milano Bicocca, Milano, Italy\label{MIBINFN} \and
Dipartimento di Fisica, Universit{\`a} degli Studi di Milano e INFN Milano,
    Milano, Italy\label{MILUINFN} \and
Institute for Nuclear Research of the Russian Academy of Sciences,
    Moscow, Russia\label{INR} \and
Institute for Theoretical and Experimental Physics,
    Moscow, Russia\label{ITEP} \and
National Research Centre ``Kurchatov Institute'', Moscow, Russia\label{KU} \and
Max-Planck-Institut f{\"ur} Physik, M{\"u}nchen, Germany\label{MPIP} \and
Physik Department and Excellence Cluster Universe,
    Technische  Universit{\"a}t M{\"u}nchen, Germany\label{TUM} \and
Dipartimento di Fisica e Astronomia dell{`}Universit{\`a} di Padova,
    Padova, Italy\label{PDUNI} \and
INFN  Padova, Padova, Italy\label{PDINFN} \and
Physikalisches Institut, Eberhard Karls Universit{\"a}t T{\"u}bingen,
    T{\"u}bingen, Germany\label{TU} \and
Physik Institut der Universit{\"a}t Z{\"u}rich, Z{\"u}rich,
    Switzerland\label{UZH}
}

\date{Received: date / Accepted: date}
\maketitle

\begin{abstract}
The \GERDA\ experiment located at the Laboratori Nazionali del Gran Sasso of
INFN searches for neutrinoless double beta (\onbb) decay of $^{76}$Ge using
germanium diodes as source and detector. In Phase~I of the experiment eight
semi-coaxial and five BEGe type detectors have been deployed.  The latter type
is used in this field of research for the first time.  All detectors are made
from material with enriched $^{76}$Ge fraction.  The experimental sensitivity
can be improved by analyzing the pulse shape of the detector signals with the
aim to reject background events.  This paper documents the algorithms
developed before the data of Phase~I were unblinded.  The double escape peak
(DEP) and Compton edge events of 2.615~MeV $\gamma$ rays from $^{208}$Tl
decays as well as two-neutrino double beta (\nnbb) decays of $^{76}$Ge are
used as proxies for \onbb\ decay.

For BEGe detectors the chosen selection is based on a single pulse shape
parameter. It accepts $0.92\pm0.02$ of signal-like events while about 80\,\%
of the background events at $Q_{\beta\beta}=2039$~keV are rejected.

For semi-coaxial detectors three analyses are developed. The one based on an
artificial neural network is used for the search of \onbb\ decay.  It retains
90\,\% of DEP events and rejects about half of the events around
$Q_{\beta\beta}$.  The \nnbb\ events have an efficiency of $0.85\pm0.02$ and
the one for \onbb\ decays is estimated to be $0.90^{+0.05}_{-0.09}$.  A second
analysis uses a likelihood approach trained on Compton edge events.  The third
approach uses two pulse shape parameters.  The latter two methods confirm the
classification of the neural network since about 90\,\% of the data events
rejected by the neural network are also removed by both of them.  In general,
the selection efficiency extracted from DEP events agrees well with those
determined from Compton edge events or from \nnbb\ decays.

\end{abstract}
\section{Introduction}
 \label{sec:intro}

The \gerda\ (GERmanium Detector Array) experiment searches for neutrinoless
double beta decay ($0\nu\beta\beta$ decay) of $^{76}$Ge.  Diodes made from
germanium with an enriched $^{76}$Ge isotope fraction serve as source and
detector of the decay.  The sensitivity to detect a signal, i.e.~a peak at the
decay's $Q$ value of 2039~keV, depends on the background level. Large efforts
went therefore into the selection of radio pure materials surrounding the
detectors.  The latter are mounted in low mass holders made from screened
copper and PTFE and are operated in liquid argon which serves as cooling
medium and as a shield against external backgrounds. The argon cryostat is
immersed in ultra pure water which provides additional shielding and vetoing
of muons by the detection of \v{C}erenkov radiation with photomultipliers. The
background level achieved with this setup is discussed in Ref.~\cite{bckgpap}.
Details of the apparatus which is located at the Laboratori Nazionali del Gran
Sasso of INFN can be found in Ref.~\cite{gerdapaper}.

It is known from past experiments that the time dependence of the detector
current pulse can be used to identify background
events~\cite{psd0,psd1,igexpsd,hdmpsd1,hdmpsd2,hdmpsd3}.  Signal events from
$0\nu\beta\beta$ decays deposit energy within a small volume if the electrons
lose little energy by bremsstrahlung (single site event, SSE).  On the
contrary, in background events from, e.g., photons interacting via multiple
Compton scattering, energy is often deposited at several locations well
separated by a few cm in the detector (multi site events, MSE).  The pulse
shapes will in general be different for the two event classes and can thus be
used to improve the sensitivity of the experiment.  Energy depositions from
$\alpha$ or $\beta$ decays near or at the detector surface lead to peculiar
pulse shapes as well that allows their identification.

\begin{figure}[t]
   \begin{center}
      \includegraphics[width=0.8\columnwidth]{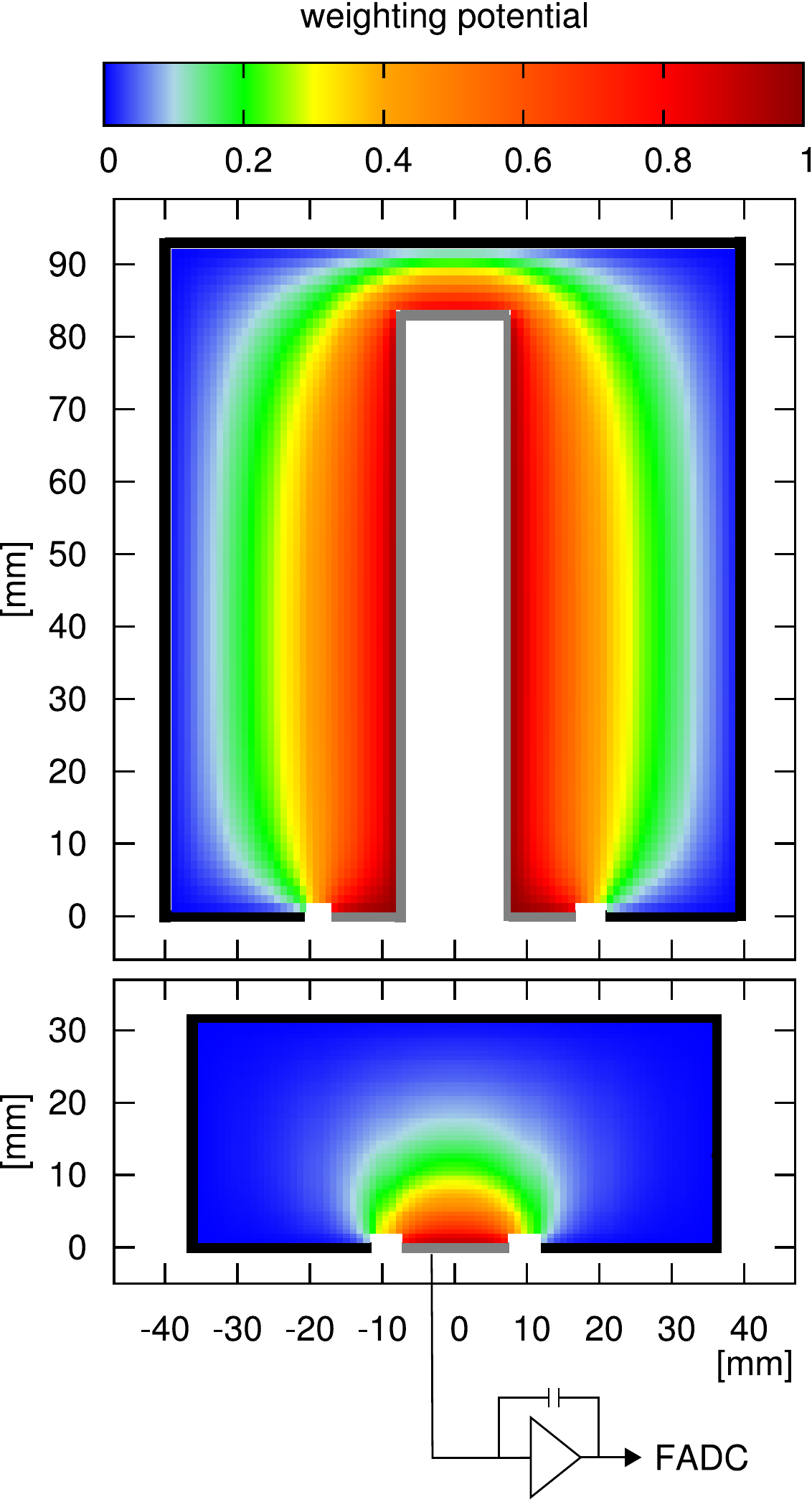}
   \end{center}
   \caption{ \label{fig:wp}
     Cross section of a semi-coaxial detector (top) and a BEGe detector
     (bottom).  The {\it p+} electrode is drawn in grey and the {\it n+}
     electrode in black (thickness not to scale).  The electrodes are
     separated by an insulating groove.  Color profiles of the weighting
     potential~\cite{ShockleyRamo} are overlayed on the detector drawings.
     Also sketched for the BEGe is the readout with a charge sensitive
     amplifier.
}
\end{figure}

\gerda\ proceeds in two phases. In Phase~I, five semi-coaxial diodes from the
former Heidelberg-Moscow (\HdM) experiment (named ANG~1 - ANG~5)~\cite{hdm}
and three from the \IGEX\ experiment (named RG~1 - RG~3)~\cite{igex} are
deployed.  For Phase~II, 30 new detectors of BEGe type~\cite{canberra} have
been produced of which five have already been deployed for part of Phase~I
(GD32B, GD32C, GD32D, GD35B and GD35C).  The characteristics of all detectors
are given in Refs.~\cite{bckgpap,gerdapaper}.

Each detector is connected to a charge sensitive amplifier and the output is
digitized with Flash ADCs with 100 MHz sampling frequency.  The deposited
energy and the parameters needed for pulse shape analysis are reconstructed
offline~\cite{gelatio1,gelatio2} from the recorded pulse.

The effect of the PSD selection on the physics data is typically always
compared in the energy interval 1930 - 2190~keV which is used for the
$0\nu\beta\beta$ analysis~\cite{bckgpap}.  The blinded energy window 2034 -
2044~keV and two intervals 2099 - 2109~keV (SEP of $^{208}$Tl line) and 2114 -
2124~keV ($^{214}$Bi line) are removed.  The remaining energy range is
referred to as the ``230~keV window'' in the following.

Events with an energy deposition in the window $Q_{\beta\beta}\pm 5$~keV
($Q_{\beta\beta}\pm4$~keV) were hidden for the semi-coaxial (BEGe) detectors
and were analyzed after all selections and calibrations had been finalized.
This article presents the pulse shape analysis for \GERDA\ Phase~I developed
in advance of the data unblinding.

\section{Pulse shape discrimination}
\label{sec:psd}

Semi-coaxial and BEGe detectors have different geometries and hence different
electric field distributions.  Fig.~\ref{fig:wp} shows a cross section of a
semi-coaxial and a BEGe detector with the corresponding weighting potential
profiles. The latter determine the induced signal on the readout electrode for
drifting charges at a given position in the diode~\cite{ShockleyRamo}.  For
both detectors, the bulk is {\it p} type, the high voltage is applied to the
{\it n+} electrode and the readout is connected to the {\it p+} electrode.
The electrodes are separated by an insulating groove.

\subsection{BEGe detectors}
  \label{ssec:psdbege}

The induced current pulse is largest when charges drift through the volume of
a large weighting potential gradient.  For BEGe detectors this is the case
when holes reach the readout electrode. Electrons do not contribute much since
they drift through a volume of low field strength.  The electric field profile
in BEGes causes holes to approach the {\it p+} electrode along very similar
trajectories, irrespective where the energy deposition
occurred~\cite{PSSpaper}. For a localized deposition consequently, the maximum
of the current pulse is nearly always directly proportional to the energy.
Only depositions in a small volume of 3-6\,\% close to the {\it p+} electrode
exhibit larger current pulse maxima since electrons also contribute in this
case~\cite{PSSpaper,PSDpaper}.  This behavior motivates the use of the ratio
$A/E$ for pulse shape discrimination (PSD) with $A$ being the maximum of the
current pulse and $E$ being the energy.  The current pulses are extracted from
the recorded charge pulses by differentiation.

\begin{figure*}[t]
   \begin{center}
     \includegraphics[width=0.9\textwidth]{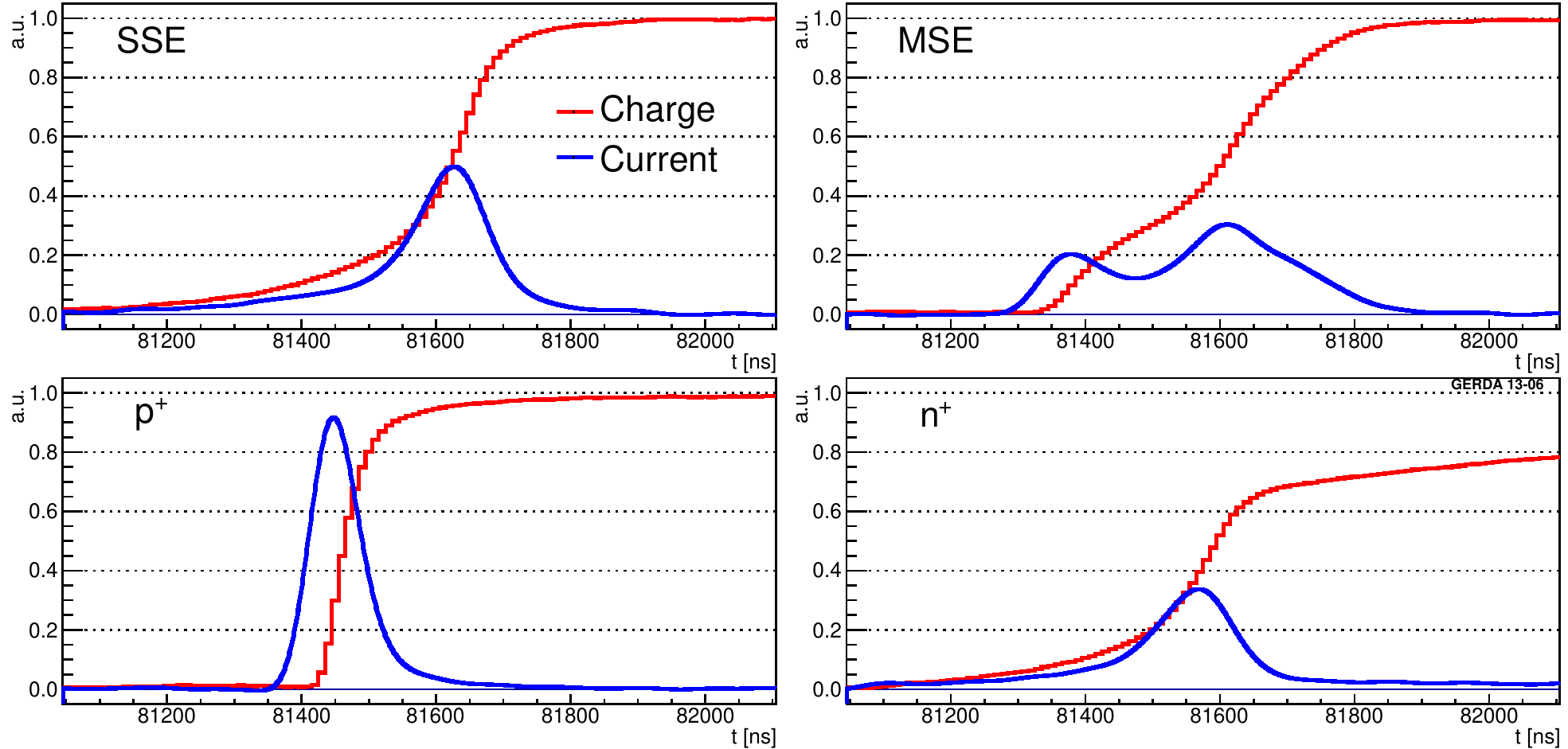}
   \caption{ \label{fig:bege_pulse}
         Candidate pulse traces taken from BEGe data for a SSE (top left), MSE
         (top right), {\it p+} electrode event (bottom left) and {\it n+}
         surface event (bottom right). The maximal charge pulse amplitudes are
         set equal to one for normalization and current pulses
         have equal integrals. The current pulses are interpolated.
}
   \end{center}
\end{figure*}

For double beta decay events ($0\nu\beta\beta$ or two-neutrino double beta
decay, $2\nu\beta\beta$), the energy is mostly deposited at one location in
the detector (SSE).  Fig.~\ref{fig:bege_pulse} (top left) shows an example of
a possible SSE charge and current trace from the data.  For SSE in the bulk
detector volume one expects a nearly Gaussian distribution of $A/E$ with a
width dominated by the noise in the readout electronics.

For MSE, e.g. from multiple Compton scattered $\gamma$ rays, the current
pulses of the charges from the different locations will have -- in general --
different drift times and hence two or more time-separated current pulses are
visible. For the same total energy $E$, the maximum current amplitude $A$ will
be smaller in this case.  Such a case is shown in the top right plot of
Fig.~\ref{fig:bege_pulse}.

For surface events near the {\it p+} electrode the current amplitude, and
consequently $A/E$, is larger and peaks earlier in time than for a standard
SSE.  This feature allows these signals to be recognized
efficiently~\cite{MatteoThesis}.  A typical event is shown in the bottom left
trace of Fig.~\ref{fig:bege_pulse}.

The {\it n+} electrode is formed by infusion of lithium, which diffuses
inwards resulting in a fast falling concentration profile starting from
saturation at the surface.  The {\it p-n} junction is below the {\it n+}
electrode surface.  Going from the junction towards the outer surface, the
electric field decreases.  The point when it reaches zero corresponds to the
edge of the conventional {\it n+} electrode dead layer, that is 0.8 - 1~mm
thick (1.5 - 2.3~mm) for the BEGe (semi-coaxial) detectors.  However, charges
(holes) from particle interactions can still be transferred from the dead
layer into the active volume via diffusion (see
e.g. Ref.~\cite{MajoranaDLstudy}) up to the point near the outer surface where
the Li concentration becomes high enough to result in a significant
recombination probability. Due to the slow nature of the diffusion compared to
the charge carrier drift in the active volume, the rise time of signals from
interactions in this region is increased.  This causes a ballistic deficit
loss in the energy reconstruction. The latter might be further reduced by
recombination of free charges near the outer surface.  The pulse integration
time for $A$ is $\sim$100 times shorter than the one for energy causing an
even stronger ballistic deficit and leading to a reduced $A/E$ ratio.  This is
utilized to identify $\beta$ particles penetrating through the {\it n+}
layer~\cite{nplusRef}.  The bottom right trace of Fig.~\ref{fig:bege_pulse}
shows a candidate event.

A pulse shape discrimination based on $A/E$ has been developed in preparation
for Phase~II. It is applied here and has been tested extensively before
through experimental measurements both with detectors operated in vacuum
cryostats~\cite{PSDpaper} and in liquid argon~\cite{BEGeLAr,LArGeBEGe,markphd}
as well as through pulse-shape simulations~\cite{PSSpaper}.

For double beta decay events, bremsstrahlung of electrons can reduce $A$ and
and results in a low side tail of the $A/E$ distribution while events close to
the {\it p+} electrode cause a tail on the high side.  Thus the PSD survival
probability of double beta decay is $<$1.

\subsection{Semi-coaxial detectors}

\begin{figure}[t]
   \begin{center}
      \includegraphics[width=0.9\columnwidth]{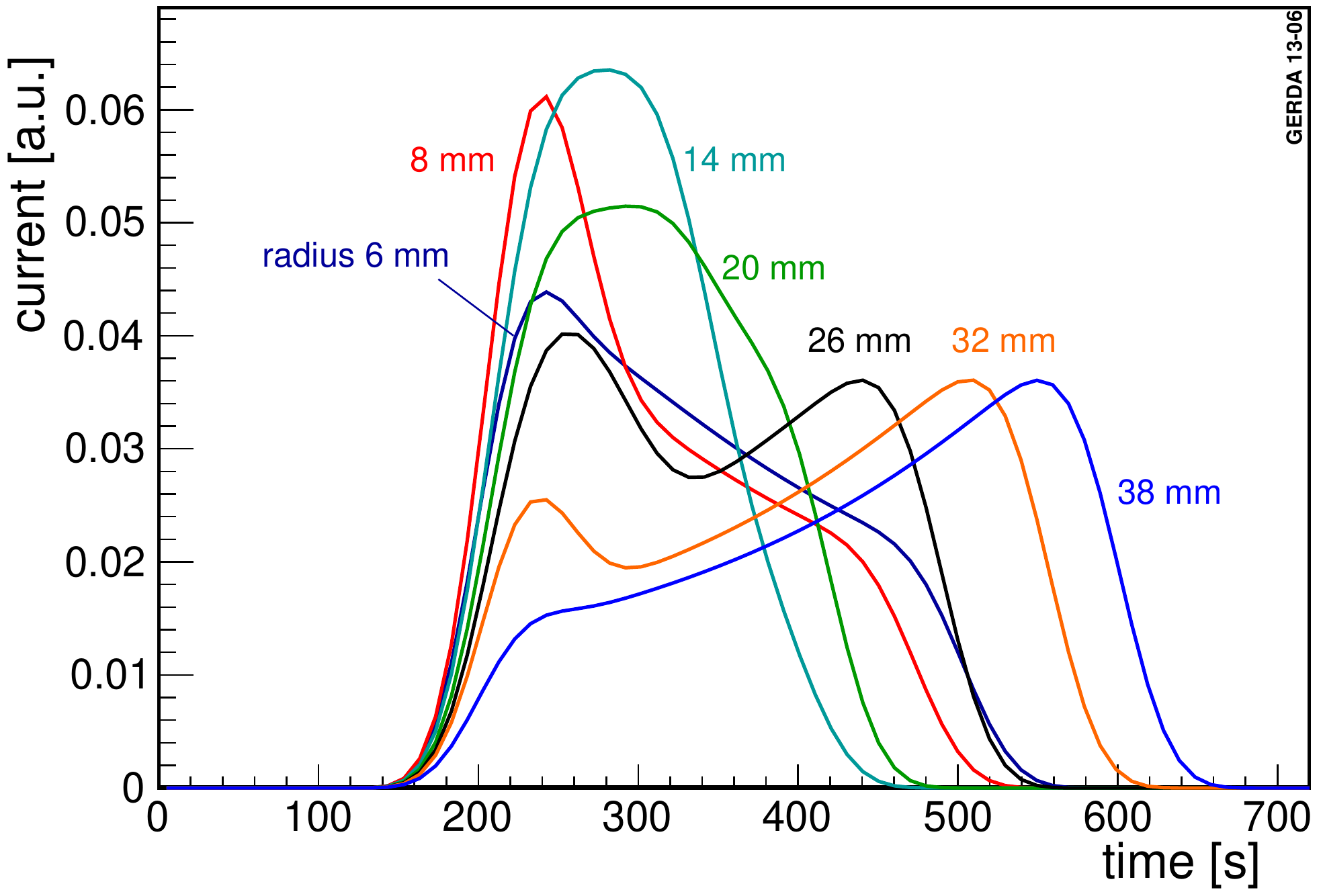}
   \caption{ \label{fig:coax_sse}
         Simulated pulse shapes for SSE in a semi-coaxial detector.  The
         locations vary from the outer {\it n+} surface (radius 38~mm) towards
         the bore hole (radius 6~mm) along a radial line at the midplane in
         the longitudinal direction.  The integrals of all pulses are the
         same. The pulses are shaped to mimic the limited bandwidth of the
         readout electronics.
}
   \end{center}
\end{figure}

For semi-coaxial detectors, the weighting field also peaks at the {\it p+}
contact but the gradient is lower and hence a larger part of the volume is
relevant for the current signal.  Fig.~\ref{fig:coax_sse} shows examples of
current pulses from localized energy depositions. These simulations have been
performed using the software described in Refs.~\cite{PSSpaper,MGS}. For
energy depositions close to the {\it n+} surface (at radius 38~mm in
Fig.~\ref{fig:coax_sse}) only holes contribute to the signal and the current
peaks at the end. In contrast, for surface {\it p+} events close to the bore
hole (at radius 6~mm) the current peaks earlier in time. This behavior is
common to BEGe detectors. Pulses in the bulk volume show a variety of
different shapes since electrons and holes contribute.  Consequently, $A/E$ by
itself is not a useful variable for coaxial detectors. Instead three
significantly different methods have been investigated. The main one uses an
artificial neural network to identify single site events; the second one
relies on a likelihood method to discriminate between SSE like events and
background events; the third is based on the correlation between $A/E$ and the
pulse asymmetry visible in Fig~\ref{fig:coax_sse}.

\subsection{Pulse shape calibration}
\label{sec:psd_calibration}

Common to all methods and for both detector types is the use of calibration
data, taken once per week, to test the performance and -- in case of pattern
recognition programs -- to train the algorithm.  The $^{228}$Th calibration
spectrum contains a peak at 2614.5~keV from the $^{208}$Tl decay. The double
escape peak (DEP, at 1592.5~keV) of this line is used as proxy for SSE while
full energy peaks (FEP, e.g. at 1620.7~keV) or the single escape peak (SEP, at
2103.5~keV) are dominantly MSE.  The disadvantage of the DEP is that the
distribution of the events is not homogeneous inside the detector as it is for
$0\nu\beta\beta$ decays. Since two 511~keV photons escape, DEP events are
dominantly located at the corners.  Events due to Compton scattering of
$\gamma$ rays span a wide energy range and also contain a large fraction of
SSE. Therefore they are also used for characterizing the PSD methods,
especially their energy dependencies.

The $2\nu\beta\beta$ decay is homogeneously distributed and thus allows a cross
check of the signal detection efficiency of the PSD methods.

\section{Pulse shape discrimination for BEGe detectors}

 BEGe detectors from Canberra~\cite{canberra} feature not only a small
 detector capacitance and hence very good energy resolution but also allow a
 superior pulse shape discrimination of background events compared to
 semi-coaxial detectors.  The PSD method and its performance is discussed in
 this section.  The full period of BEGe data taking during Phase~I (July 2012
 - May 2013) with an exposure of 2.4~kg$\cdot$yr is used in this analysis.
 One of the five detectors (GD35C) was unstable and is not included in the
 data set.

\subsection{PSD calibration}
\label{ssec:calibration}

\begin{figure}
   \begin{center}
	\includegraphics[width=0.95\columnwidth]{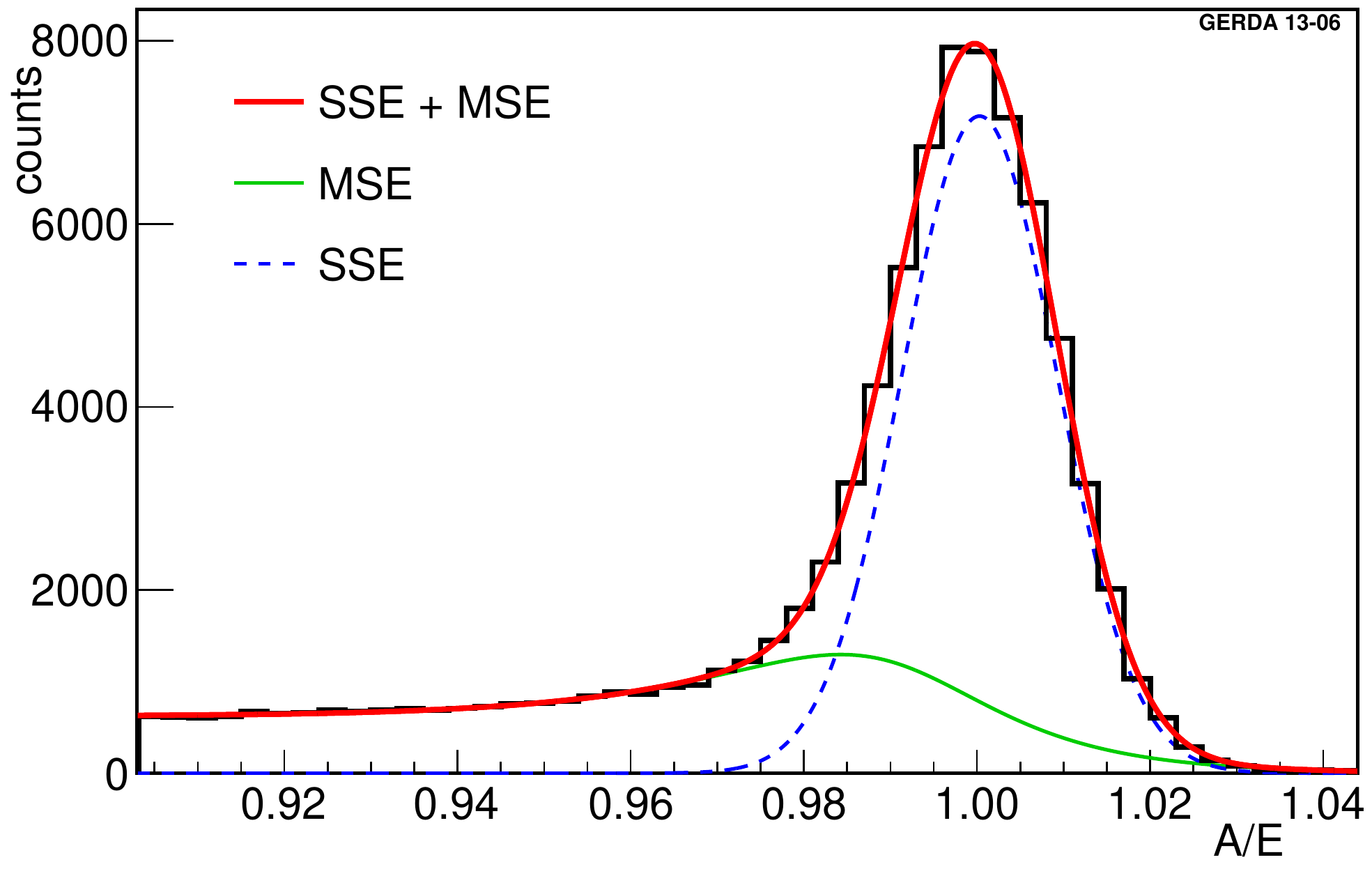}
   \caption{ \label{fig:ae_function}
           $A/E$ distribution for Compton continuum data fitted with
           function~\eqref{eq:AE}.  The dashed blue
           curve is the Gaussian component
           and the green curve is the component approximating the MSE
           contribution.
   }
   \end{center}
\end{figure}

Compton continuum and DEP events from $^{228}$Th calibration and the events in
the $2\nu\beta\beta$ energy range in physics data feature $A/E$ distributions
with a Gaussian part from SSE and a low side tail from MSE as shown in
Fig.~\ref{fig:ae_function}. It can be fitted by the function:
\begin{eqnarray}
f(x=A/E) & = &\frac{n}{\sigma_{A/E} \cdot \sqrt{2 \pi}} \cdot
                              e^{-\frac{(x - \mu_{A/E})^2}{2\sigma_{A/E}^2} } \nonumber \\
         & & \,\,  +\,\,\, m \cdot \frac{e^{ f \cdot (x - l) } +d}{e^{ (x - l)/t} +l}
    \label{eq:AE}
\end{eqnarray}
where the Gaussian term is defined by its mean $\mu_{A/E}$, standard deviation
$\sigma_{A/E}$ and integral $n$. The MSE term is parameterized empirically by
the parameters $m$, $d$, $f$, $l$ and $t$.  $\sigma_{A/E}$ is dominated by the
resolution $\sigma_A$ of $A$ which is independent of the energy, i.e.~for low
energies $\sigma_{A/E} \propto \sigma_A/E \propto 1/E $.

\begin{figure}[b]
   \begin{center}
      \includegraphics[width=0.95\columnwidth]{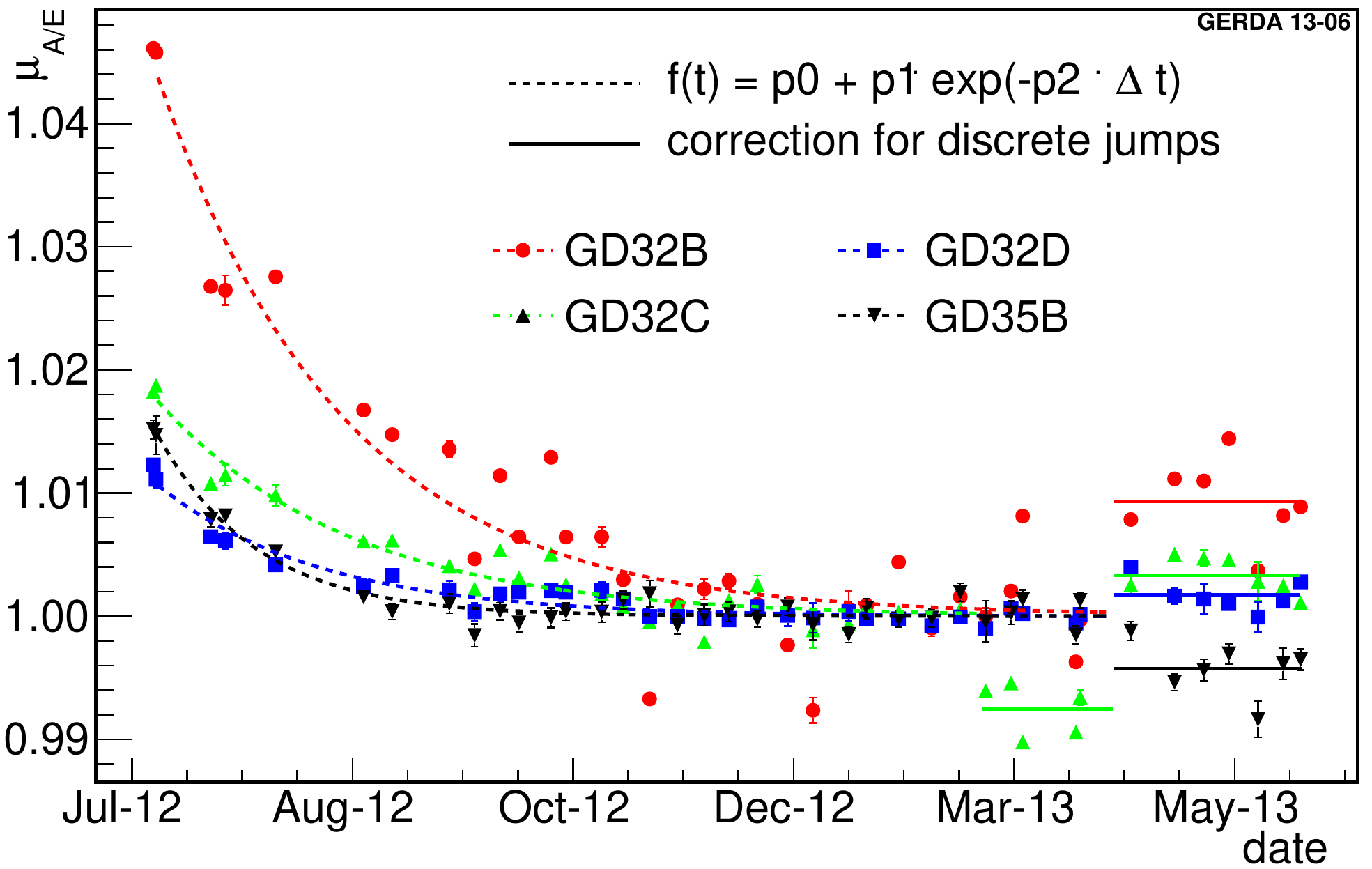}
   \caption{ \label{fig:ae_time}
           Gaussian mean $\mu_{A/E}$ for DEP events for individual $^{228}$Th
           calibrations.  The data points in the period before the occurrence
           of jumps are fitted with an exponential function as specified.
           Each $A/E$ distribution is normalized such that the
           constant of the fit ($p0$) is one.  Separate constant corrections
           are determined as averages over the periods corresponding to the
           discrete jumps.
}
   \end{center}
\end{figure}
\begin{figure}[b] 
   \begin{center}
      \includegraphics[width=0.9\columnwidth]{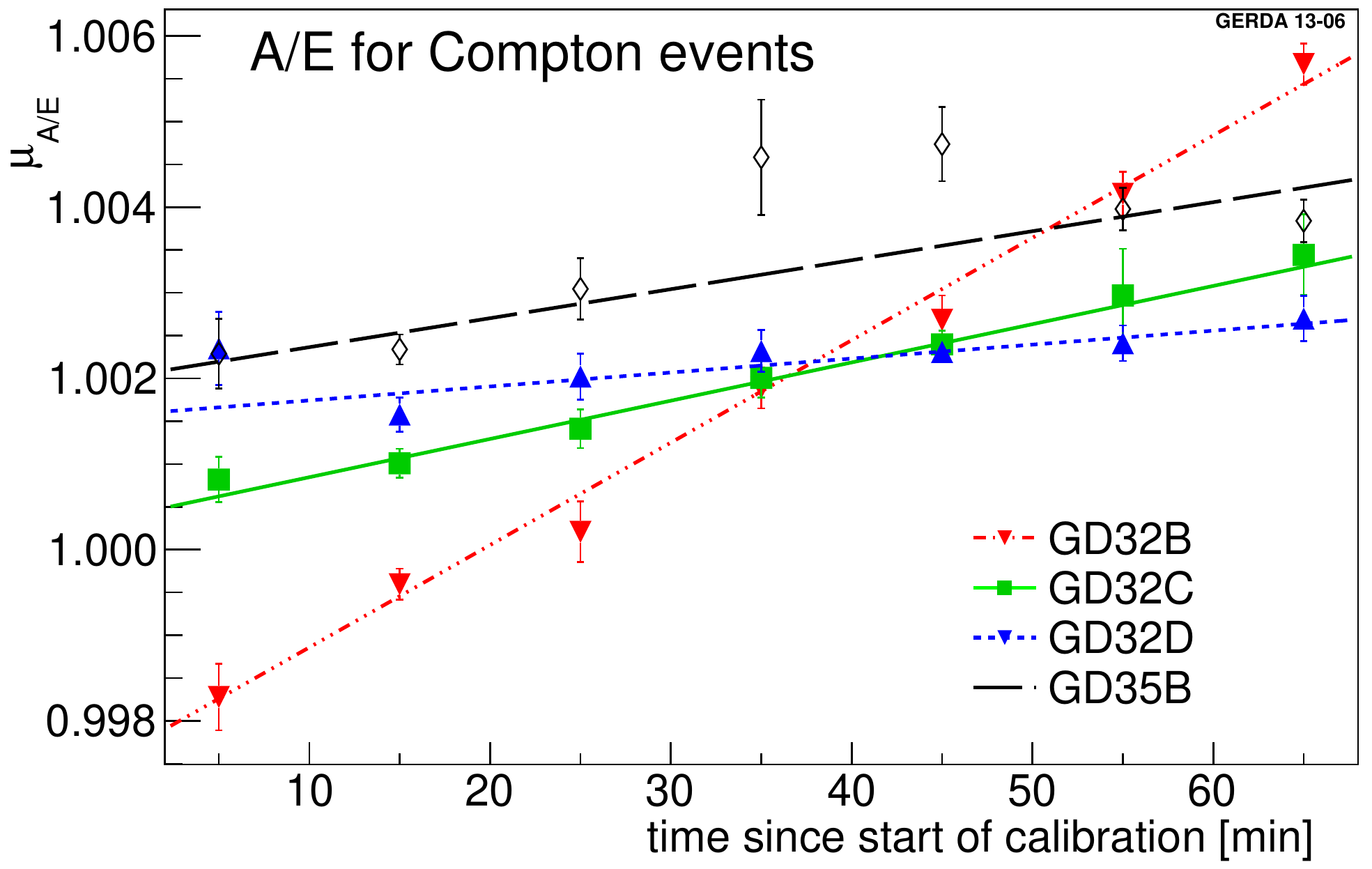}
   \caption{ \label{fig:ae_time_calib}
          Gaussian mean $\mu_{A/E}$ of the $A/E$ distribution for Compton
          events as a function of the time since the start of a calibration
          run.  The data from all calibrations are combined after the
          correction according to Fig.~\ref{fig:ae_time} has been applied.
}
   \end{center}
\end{figure}

There are a few effects which are corrected in the order they are discussed
below.  To judge their relevance, already here it is stated that events in the
interval $0.965 < A/E < 1.07$ are accepted as signal (see
Sect.~\ref{ssec:bckgcut}).
\begin{enumerate}
\item After the deployment in July 2012, $\mu_{A/E}$ drifted with a time scale
  of about one month for all detectors (see Fig.~\ref{fig:ae_time}).  The
  total change was 1 to 5\,\% depending on the detector.  The behavior is
  fitted with an exponential function which is then used to correct $A/E$ of
  calibration and physics data as a function of time.  Additionally, jumps
  occurred e.g.~after a power failure. These are also corrected.
\item $\mu_{A/E}$ increases by up to 1\,\% during calibration runs which last
  typically one hour (Fig.~\ref{fig:ae_time_calib}).  During physics data
  taking, $\mu_{A/E}$ returns to the value from before the calibration on a
  time scale of less than 24~hours, which is short compared to the one week
  interval between calibrations.  This causes $\mu_{A/E}$ in calibrations to
  be shifted to slightly higher values compared to physics data taking.  This
  effect is largely removed by applying a linear correction in time (fit shown
  in Fig.~\ref{fig:ae_time_calib}) to calibration data.  Afterwards,
  $\mu_{A/E}$ of physics data in the interval 1.0 - 1.3~MeV agrees
  approximately with Compton events from calibration data in the same energy
  region (see Fig.~\ref{fig:ae_calib_2nbb}).
\begin{figure}[htbp]
   \begin{center}
      \includegraphics[width=0.95\columnwidth]{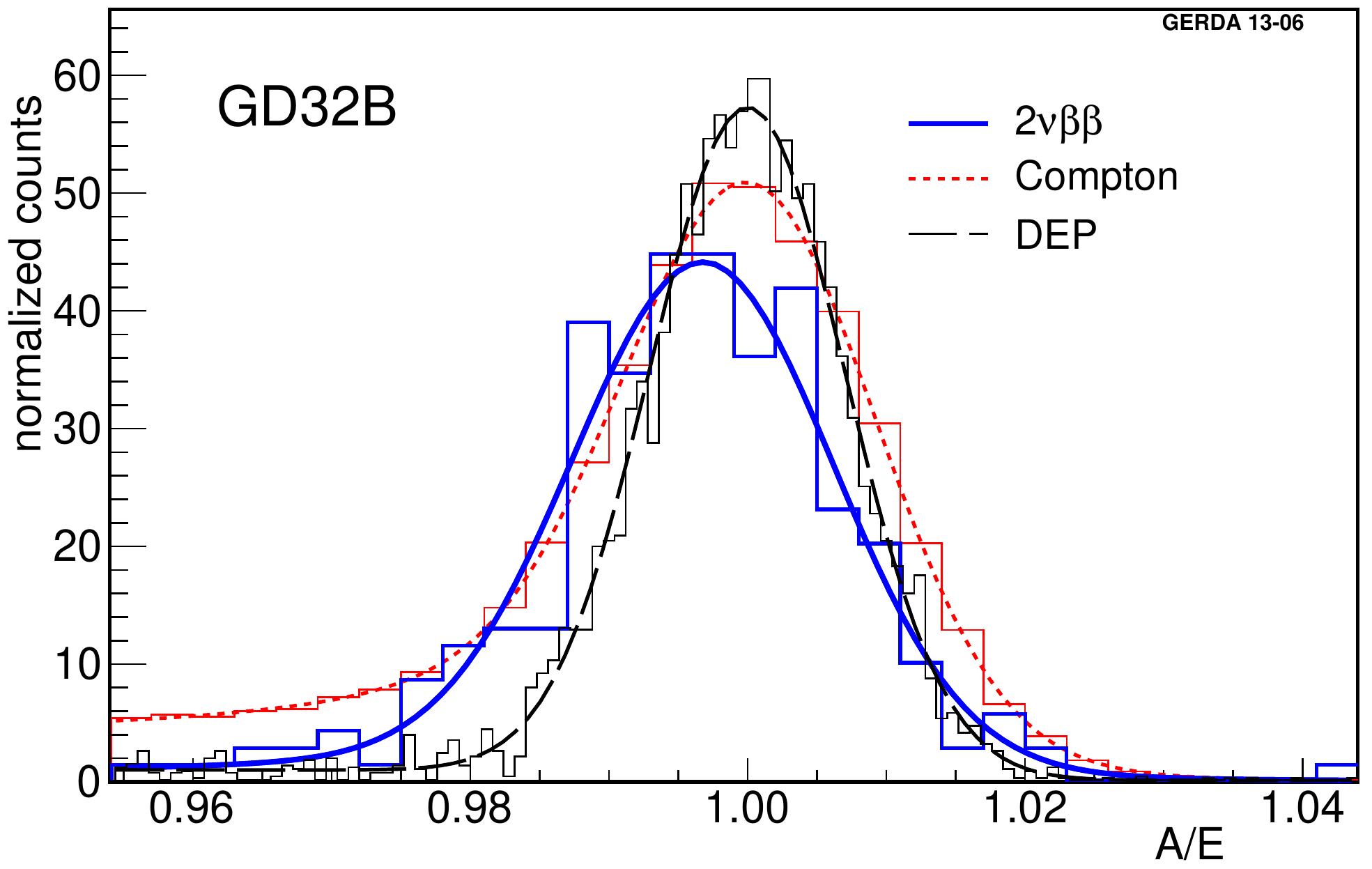}
   \caption{ \label{fig:ae_calib_2nbb}
            $A/E$ distribution of GD32B from physics data events between 1.0 and
            1.3~MeV (blue, dominantly $2\nu\beta\beta$ decays), Compton
            continuum in the same energy range (red) and DEP events
            (black). The latter two are taken from the sum of all calibrations.
            All corrections are applied.
            The tail on the left side of the Gaussian is larger in the Compton
            events due to a higher fraction of MSE compared to the physics data
            in this energy range.
}
   \end{center}
\end{figure}

\begin{figure}[b]
   \begin{center}
      \includegraphics[width=0.95\columnwidth]{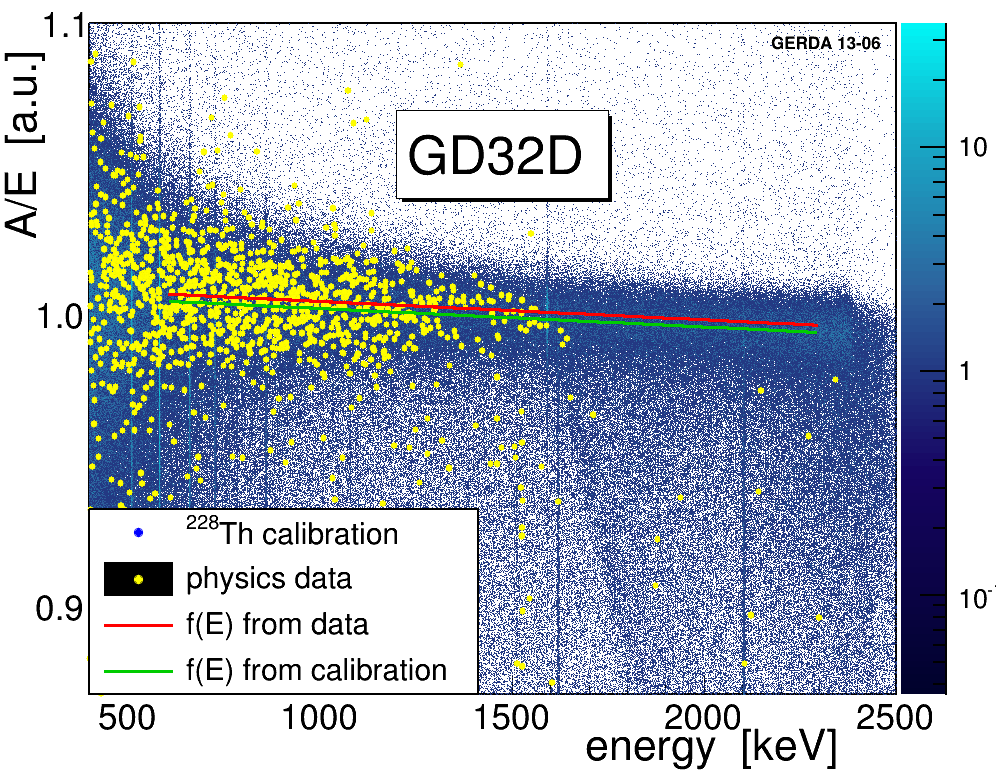}
   \caption{ \label{fig:ae(E)_Heng-Ye}
             $A/E$ energy dependence shown with $^{228}$Th calibration data
             (blue density plot) and events from physics data taking
             (predominantly $2\nu\beta\beta$, yellow points). The
             distributions of $\mu_{A/E}$ for the different energy bins are
             fitted with a linear function (green line). The $2\nu\beta\beta$
             continuum is fitted with the same function, leaving only the
             constant of the fit free 
              (red line). The data from GD32D are shown.
}
   \end{center}
\end{figure}
\item $A/E$ shows a small energy dependence (Fig.~\ref{fig:ae(E)_Heng-Ye}).
  It is measured by determining the Gaussian mean $\mu_{A/E}$ at different
  energies in the $^{208}$Tl Compton continuum between 600 and 2300~keV.
  The size is about 0.5 to 1\,\% per MeV.  This approach is documented and
  validated in Refs.~\cite{PSDpaper,DusanThesis}.  The correction is applied to
  both calibration and physics data.
\end{enumerate}

\begin{table*}[htb]
\begin{center}
   \caption{   \label{tab:DEP-2nbb}
      Comparison of $A/E$ Gaussian mean $\mu_{A/E}$ and width $\sigma_{A/E}$
      from physics data (events between 1.0~MeV and 1.3~MeV, dominantly
      $2\nu\beta\beta$ decays) and calibration data (Compton continuum in the
      region 1.0~MeV - 1.3~MeV and DEP at 1592.5~keV) after applying all
      corrections.
}
   \begin{tabular}{lcccc}\hline
   detector & $\mu_{A/E}(2\nu\beta\beta)$ - $\mu_{A/E}$(DEP) & 
   $\mu_{A/E}(2\nu\beta\beta)$ - $\mu_{A/E}$(Compton)  & $\sigma_{A/E}(2\nu\beta\beta)$ & $\sigma_{A/E}$(Compton)\\
   \hline
    GD32B   & $-0.0032 \pm 0.0007$    & $-0.0037 \pm 0.0007$   & $0.0094 
\pm 0.0006$   & $0.0089 \pm 0.0001$\\
    GD32C   & $-0.0001 \pm 0.0011$    & $\,\,\,\,\,0.0003 \pm 0.0011$   
& $0.0096 \pm 0.0005$   & $0.0094 \pm 0.0001$\\
    GD32D   & $-0.0002 \pm 0.0009$   & $\,\,\,\,\,0.0004 \pm 0.0009$   & 
$0.0118 \pm 0.0006$   & $0.0095 \pm 0.0001$\\
    GD35B   & $\,\,\,\,\,0.0014 \pm 0.0007$   & $\,\,\,\,\,0.0018 \pm 
0.0008$   & $0.0097 \pm 0.0006$   & $0.0109 \pm 0.0001$\\ \hline
   \end{tabular}
\end{center}
\end{table*}

The corrections discussed above are empirical and result in energy and time
independent $A/E$ distributions.  The origin of the time drifts might be due
to electric charges collected from LAr on the surface of the insulating
groove.  This is a known phenomenon~\cite{grooveCharges} and pulse shape
simulations show that $A/E$ changes of the observed size are conceivable.  The
small observed energy dependence of $A/E$ (item~3) is thought to be an
artefact of data acquisition and/or signal processing.

Since $A/E$ has arbitrary units, it is convenient to rescale the distribution
at the end such that the mean of the Gaussian is unity after all
corrections. This eases the combination of all detectors.

The compatibility of calibration data with physics data after the application
of all corrections is verified in Fig.~\ref{fig:ae_calib_2nbb}.  The $A/E$
Gaussian parameters are quantitatively compared in Table~\ref{tab:DEP-2nbb}.
The agreement of $\mu_{A/E}$ for DEP and $2\nu\beta\beta$ events validates
also the energy dependence correction (item~3). Small differences remain due
to imperfections of the applied corrections. They will be taken into account
as a systematic uncertainty in the determination of the $0\nu\beta\beta$
efficiency in Sect.~\ref{ssec:acceptance}.

\begin{figure}[b]  
   \begin{center}
      \includegraphics[width=0.9\columnwidth]{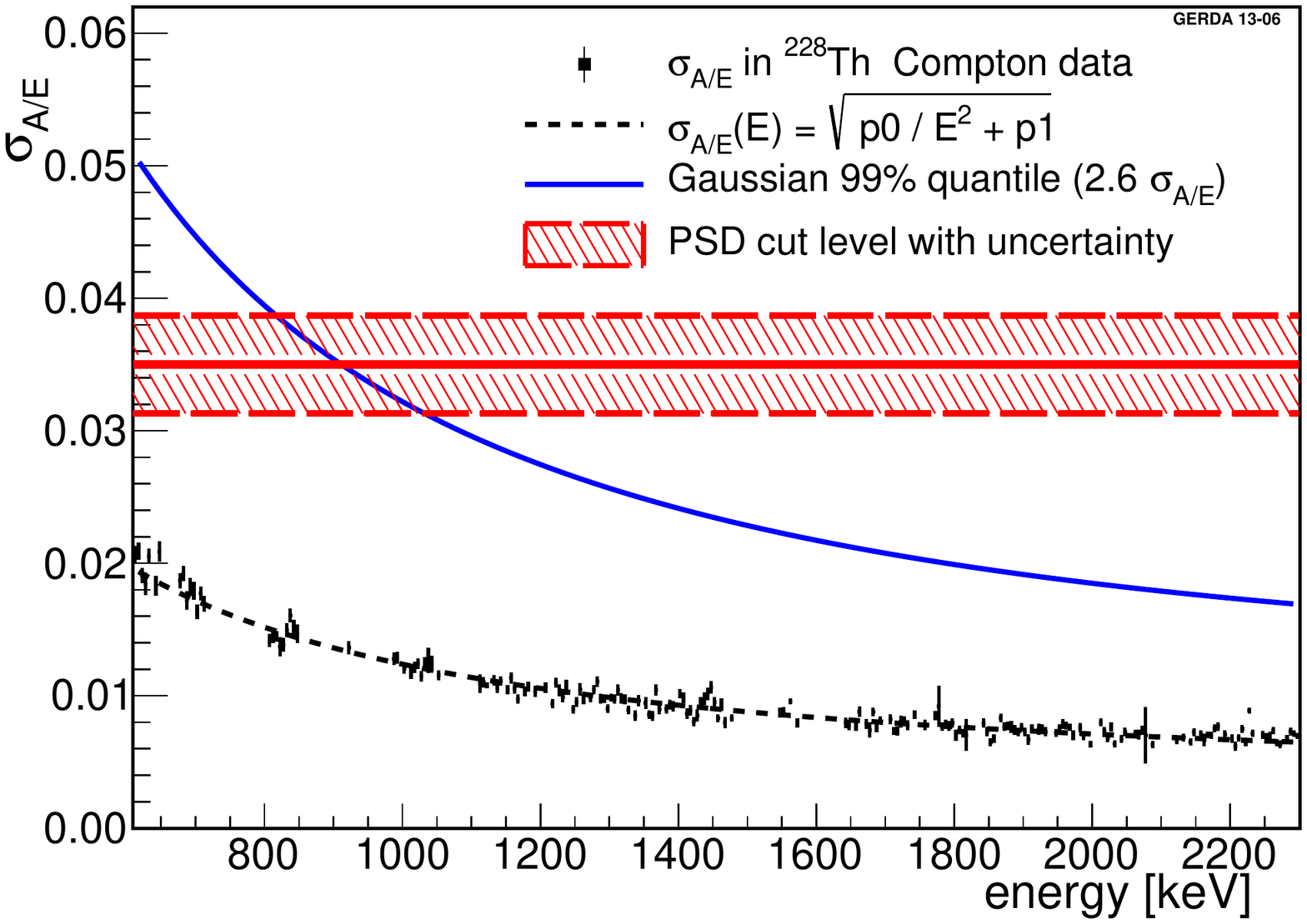}
   \caption{ \label{fig:aesigma_Heng-Ye}
         Width $\sigma_{A/E}$ of the $A/E$ Gaussian versus energy (points with
         error bars) for GD35B with a fit (black dashed line).  The blue full
         line shows the 99\,\% quantile of the Gaussian (2.6 $\sigma_{A/E}$).
         The red horizontal line corresponds to the low side PSD cut distance
         from the nominal $\mu_{A/E}=1$.  The uncertainty band is given by the
         maximal deviation of the $A/E$ scale as determined in
         Table~\ref{tab:DEP-2nbb}.
}
   \end{center}
\end{figure}

In contrast to the SSE Gaussian, the MSE part of the $A/E$ distribution and
the part from {\it p+} electrode events is only negligibly affected by the
$A/E$ resolution and its change with energy.  This motivates the use of an
$A/E$ cut that is constant at all energies: If the cut position is many
$\sigma_{A/E}$ of the Gaussian resolution away from one, the survival fraction
is practically independent of the energy.  Only at low energies this is no
longer the case.  At about 1~MeV, the cut position $A/E>0.965$ corresponds to
a separation from one by 2.6~$\sigma_{A/E}$ corresponding to the 99\,\%
quantile of a Gaussian (see Fig.~\ref{fig:aesigma_Heng-Ye}).  For lower
energies the efficiency loss of the Gaussian peak becomes relevant. Therefore
the efficiency determination is restricted to energies above 1~MeV.

The energy dependence of $\mu_{A/E}$ is determined between 600~keV and
2300~keV. Since the dependence is weak, even beyond these limits the cut
determination is accurate to within a few percent.  This is acceptable for
example to determine the fraction of $\alpha$ events at the {\it p+} electrode
passing the SSE selection cut.

\begin{figure}[b]
   \begin{center}
      \includegraphics[width=\columnwidth]{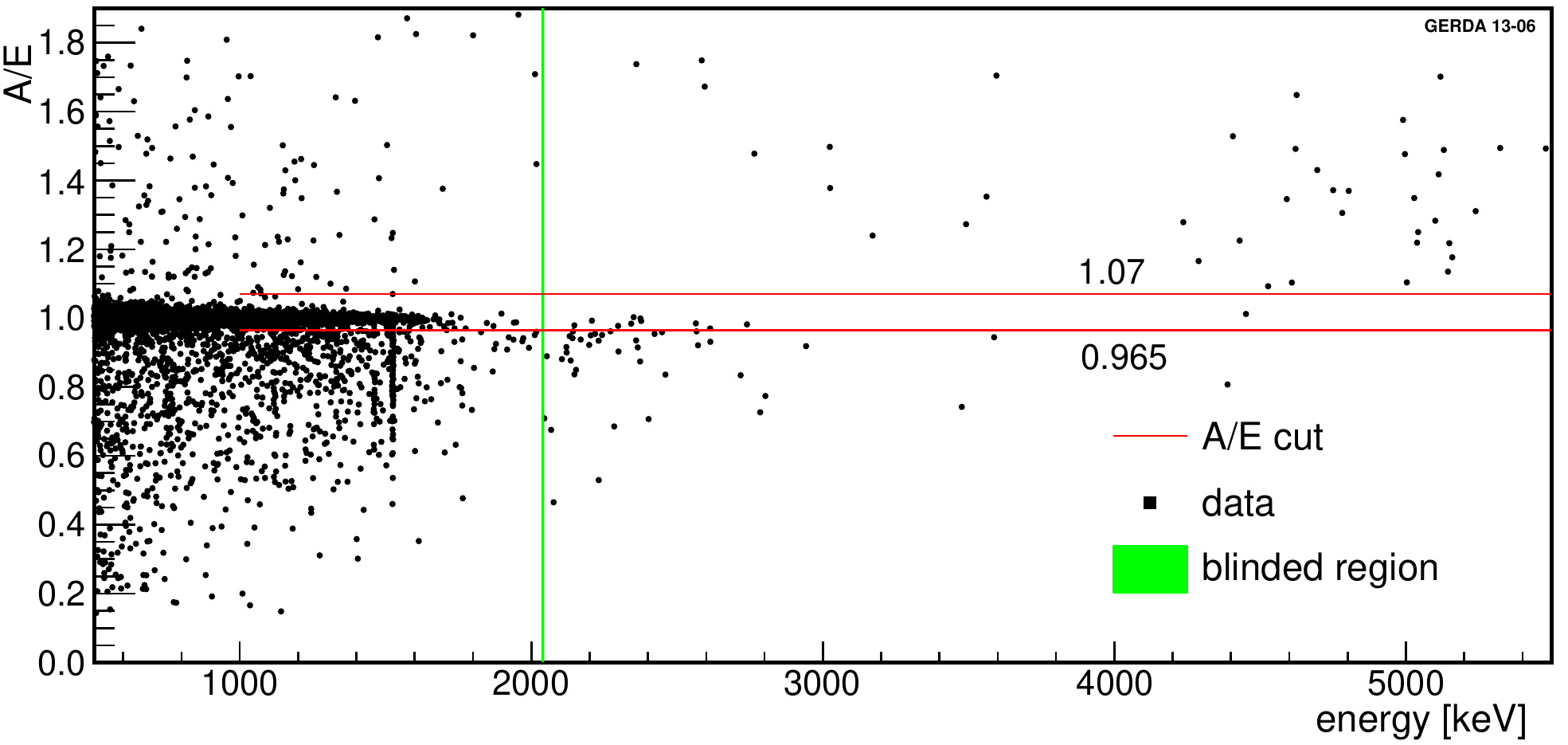}
   \caption{ \label{fig:bkgAoverE}
          $A/E$  versus energy in a wide energy range for the
          combined BEGe data set. The acceptance region boundaries are marked by
          the red lines. The blinded region is indicated by the green band.
}
   \end{center}
\end{figure}

\begin{table*}
\begin{center}
   \caption{   \label{tab:calib-cut}
       Removed fractions by the low $A/E$ cut and high $A/E$ cut and total
       surviving fractions applying both cuts in several energy regions in
       physics data and $^{228}$Th calibration data (combined data sets of all
       detectors). In the physics data set, the 1839 keV - 2239 keV region
       excludes the blinded 8~keV window around $Q_{\beta\beta}$. Peak regions
       have the underlying Compton continuum subtracted. Uncertainties are
       statistical only.
}
   \begin{tabular}[tbp]{lccc}\hline
   region & low $A/E$ cut  &  high $A/E$ cut & surviving fraction\\
	&  $A/E<0.965$ &  $A/E>1.07$ &  $0.965<A/E<1.07$\\
\hline
   $^{228}$Th calibration & & &\\
\hline
  DEP 1592.5 keV & $0.054 \pm 0.003$ & $0.015 \pm 0.001$ & $0.931 \pm 0.003$  \\
  FEP 1620.7 keV & $0.771 \pm 0.008$ & $0.009 \pm 0.002$ & $0.220 \pm 0.008$  \\
  SEP 2103.5 keV & $0.825 \pm 0.005$ & $0.011 \pm 0.001$ & $0.165 \pm 0.005$ \\
\hline
   physics data & & &\\
\hline
  FEP 1524.7 keV & $0.69 \pm 0.05$ & $0.027 \pm 0.015$ & $0.29 \pm 0.05$  \\
 1000 - 1450 keV & $0.230 \pm 0.011$ & $0.022 \pm 0.004$ & $0.748 \pm 0.011$  \\
 1839 - 2239 keV & $30/40$ & $3/40$ & $7/40$  \\
 $>$~4 MeV ($\alpha$ at {\it p+})  & $1/35$ & $33/35$ & $1/35$  \\ \hline
   \end{tabular}
\end{center}
\end{table*}

\subsection{Application of PSD to data}
\label{ssec:bckgcut}

Fig.~\ref{fig:bkgAoverE} shows $A/E$ plotted versus energy for physics data in
a wide energy range together with the acceptance range.  The data of all
detectors have been added after all applicable corrections and the
normalization of the Gaussian mean to one.  The cut rejects events with
$A/E<0.965$ (``low $A/E$ cut'') or $A/E>1.07$ (``high $A/E$ cut'').  The high
side cut interval was chosen twice wider due to the much lower occurrence and
better separation of {\it p+} electrode events.  The cut levels result in a
high probability to observe no background event in the final $Q_{\beta\beta}$
analysis window for the Phase~I BEGe data set, while maintaining a large
efficiency with small uncertainties. As can be seen from
Fig.~\ref{fig:aesigma_Heng-Ye}, at Q$_{\beta\beta}$ the cut is $\geq
4.5$~$\sigma_{A/E}$ apart from one.

\begin{figure}[b]
   \begin{center}
      \includegraphics[width=\columnwidth]{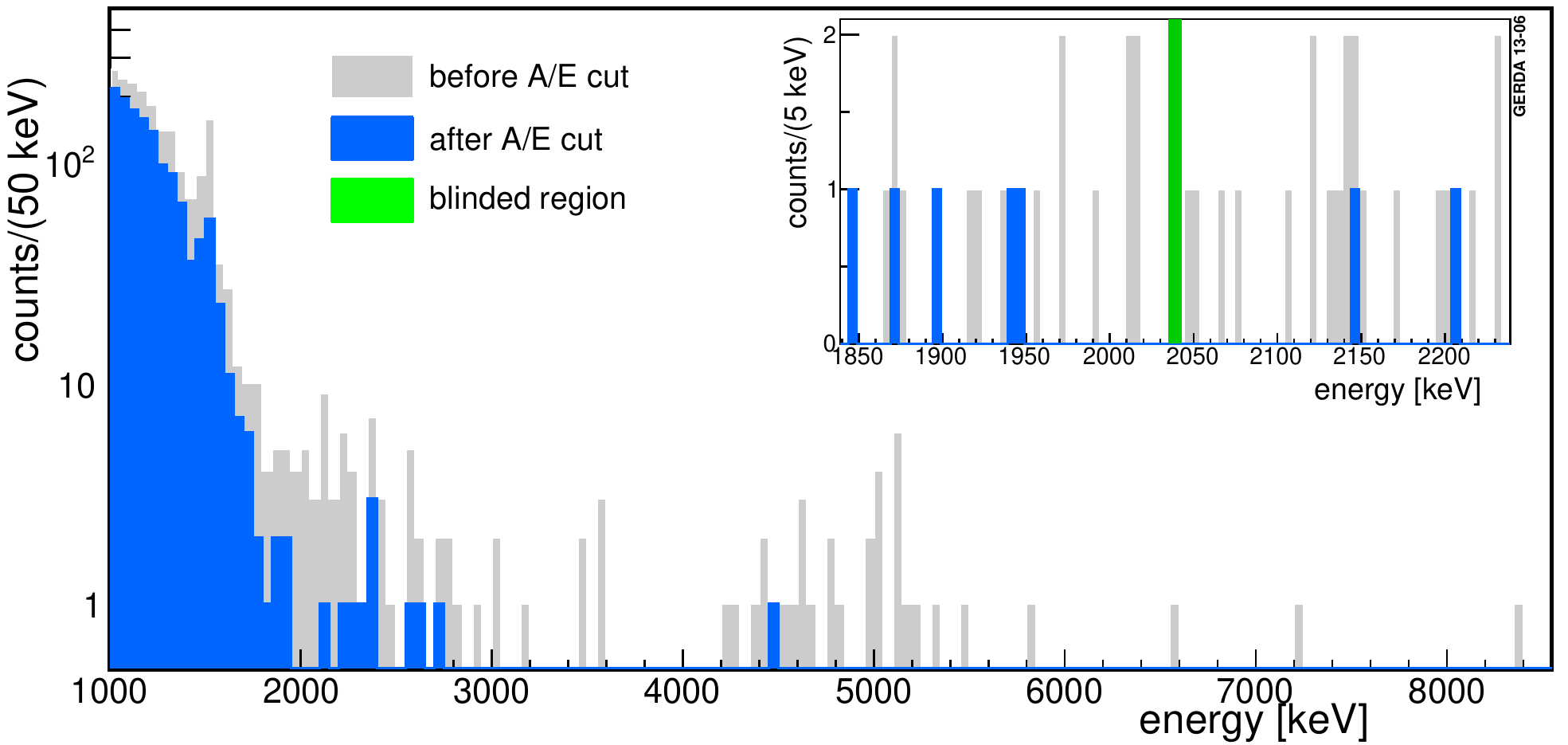}
   \caption{ \label{fig:bkg-spectrum}
          Energy spectrum of the combined BEGe data set: grey (blue)  before
          (after) the PSD cut. The inset shows a zoom at the region
          Q$_{\beta\beta} \pm 200$~keV with the 8~keV blinded region  in green.
}
   \end{center}
\end{figure}

Fig.~\ref{fig:bkg-spectrum} shows the combined energy spectrum of the BEGe
detectors before and after the PSD cut. In the physics data set with
2.4~kg$\cdot$yr exposure, seven out of 40 events in the 400~keV wide region
around $Q_{\beta\beta}$ (excluding an 8~keV blinding window) are kept and
hence the background for BEGe detectors is reduced from (0.042~$\pm$~0.007) to
(0.007$_{-0.002}^{+0.004}$)~\ctsper. In the smaller 230~keV region three out
of 23 events remain.  Table~\ref{tab:calib-cut} shows the surviving fractions
for several interesting energy regions in the physics data and $^{228}$Th
calibration data.  The suppression of the $^{42}$K $\gamma$ line at 1525~keV
in physics data is consistent with the one of the $^{212}$Bi line at 1621~keV.
The rejection of $\alpha$ events at the {\it p+} electrode is consistent with
measurements with an $\alpha$ source in a dedicated setup~\cite{MatteoThesis}.

The energy spectrum of the physics data can be used to identify the background
components at $Q_{\beta\beta}$ as described in Ref.~\cite{bckgpap}.  About
half of the events are from $^{42}$K decays on the {\it n+} electrode surface
which are rejected by the low side $A/E$ cut with large
efficiency~\cite{nplusRef}.  About one third of the background at
$Q_{\beta\beta}$ is due to $^{214}$Bi and $^{208}$Tl. Their survival
probability can be determined from the calibration data (52\,\% for
$^{208}$Tl) or extrapolated from previous studies~\cite{LArGeBEGe,markphd}
(36\,\% for $^{214}$Bi).  The remaining backgrounds e.g.~from $^{68}$Ga inside
the detectors and from the {\it p+} surface are suppressed
efficiently~\cite{PSSpaper,MatteoThesis}. The rejection of 80\,\% of the
physics events at $Q_{\beta\beta}$ is hence consistent with expectation.

\begin{figure}[b]
   \begin{center}
      \includegraphics[width=0.93\columnwidth]{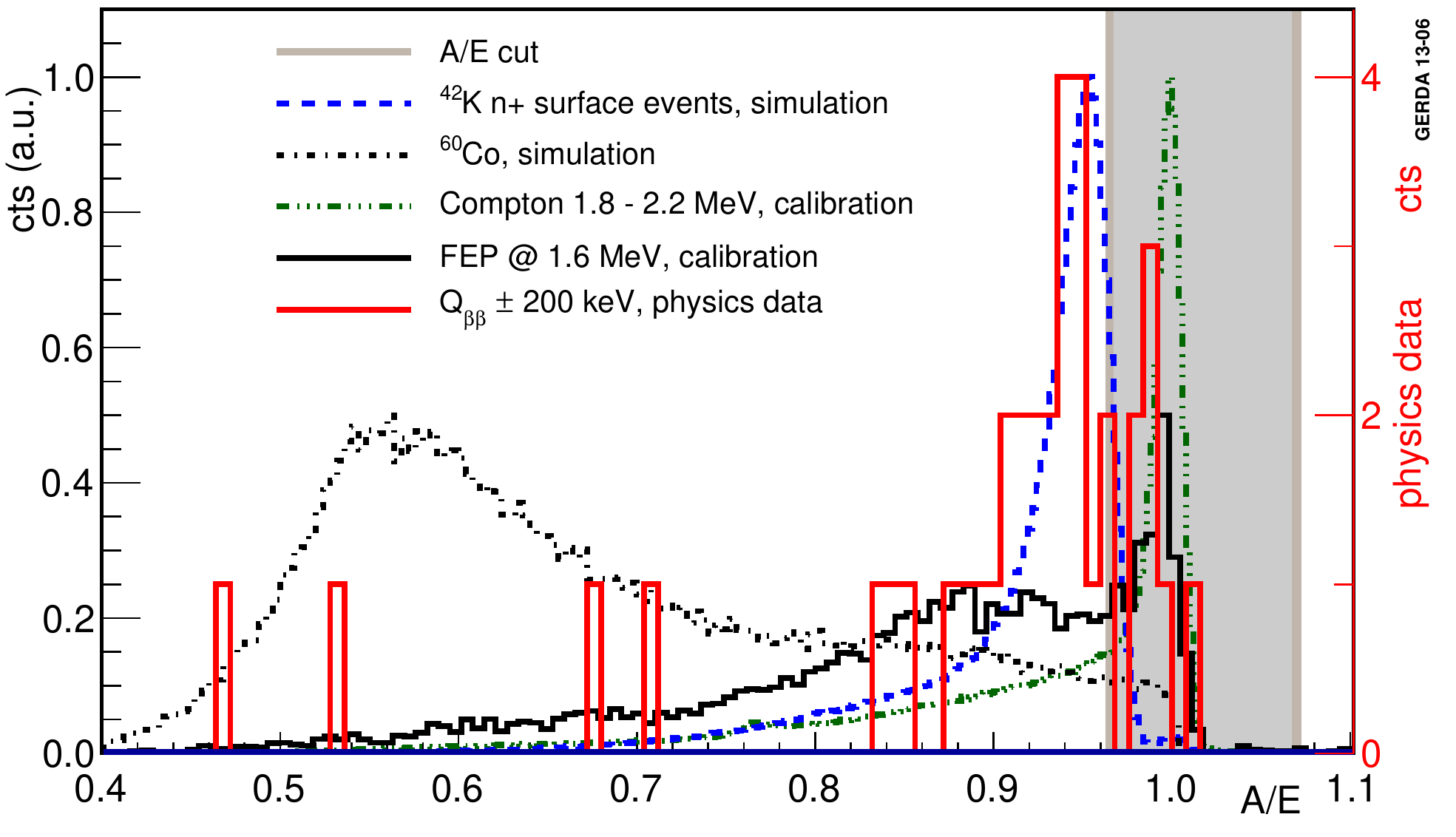}
   \caption{ \label{fig:Mix}
           $A/E$ histogram of the physics data within 200~keV of \qbb\ (red)
           compared to Compton continuum events (green dot-dot-dashed) and
           1621~keV FEP events (black) from calibration data. Also shown are
           simulations of $^{42}$K decays at the {\it n+} electrode surface
           (blue dashed) and $^{60}$Co (black dot-dashed)~\cite{PSSpaper}. The
           scalings of the histograms are arbitrary. Three physics data events
           have large $A/E$ values ({\it p+} electrode events) and are out of
           scale. The accepted interval is shown in grey.
}
   \end{center}
\end{figure}
In Fig.~\ref{fig:Mix}, the $A/E$ distribution of physics data in the
Q$_{\beta\beta} \pm 200$~keV region is compared with the distributions from
different background sources.  The peak at 0.94 can be attributed to {\it n+}
surface events.  The $A/E$ distribution of the other events is compatible
within statistical uncertainty with the ones expected from the different
background sources.

\subsection{Evaluation of $0\nu\beta\beta$ cut survival fraction for BEGes}
\label{ssec:acceptance}

The PSD survival fraction of DEP events can vary from the one for
$0\nu\beta\beta$ events because of the difference of the event locations in a
detector (see Sect.~\ref{sec:psd_calibration}) and due to the different
energy release and the resulting bremsstrahlung emission.

The influence of these effects was studied by simulations. The first effect
was irrelevant in past publications since only a low $A/E$ cut was studied and
{\it p+} electrode events have higher $A/E$.  In the present analysis, we
required also $A/E<1.07$.  Therefore we use a pulse shape simulation of
$0\nu\beta\beta$ events~\cite{PSSpaper} to determine the rejected fraction of
signal events by the high $A/E$ cut.

The second effect can influence the low $A/E$ cut survival.  To estimate its
size, we compare the pulse shape simulation result~\cite{PSSpaper} with a
Monte Carlo simulation~\cite{PSDpaper} which selects events according to the
bremsstrahlung energy.  The latter is approximately equivalent to a cut on the
spatial extent of the interaction since higher energy bremsstrahlung $\gamma$
rays interact farther from the main interaction site (electron-positron pair
creation vertex for DEP or $0\nu\beta\beta$ decay vertex). The fraction of DEP
events with a Compton scattering before the pair creation was taken into
account. The determined fraction of MSE in DEP and $0\nu\beta\beta$ events was
the same within uncertainties.  In contrast, the pulse shape simulation
removes 1.8\,\% events more for $A/E<0.965$.  This difference could be caused
by a larger fraction of bremsstrahlung in $0\nu\beta\beta$ compared to DEP or
due to simulation artefacts~\cite{PSSpaper}. Here we follow the result of the
Monte Carlo simulation, i.e.~use the DEP survival fraction for the low $A/E$
cut, and take the difference to the pulse shape simulation as systematic
error.

Thus, the survival fraction $\epsilon_{0\nu\beta\beta}$ of the $0\nu\beta\beta$
signal is estimated as follows:
\begin{itemize}
\item the rejected fraction for the low side cut of 0.054 is determined
       from DEP events (Table~\ref{tab:calib-cut}). 
     This value varies from 0.042~$\pm$~0.006 to 0.062~$\pm$~0.010
     for the different detectors and is hence within uncertainties the
     same for all of them.  
\item the rejected fraction by the high $A/E$ cut of 0.025 is 
       determined from the $0\nu\beta\beta$
     pulse-shape simulation~\cite{PSSpaper}.
\end{itemize}

Finally, the efficiency is $\epsilon_{0\nu\beta\beta} =0.92\pm 0.02$.
The uncertainty is the quadratic sum of the following components:
\begin{itemize}
\item statistical uncertainty of the DEP survival fraction: 0.003
\item uncertainty from the $A/E$ energy dependence (item~3 in
  Sect.~\ref{ssec:calibration}): 7.5~$\cdot$~$10^{-5}$
\item uncertainty due to the residual differences between calibration and
  physics data (change of the cut by the largest difference between
  $\mu_{A/E}$ for $2\nu\beta\beta$ and Compton events in
  Table~\ref{tab:DEP-2nbb}): 0.004
\item systematic uncertainty due to the difference between the survival
    fraction of  $0\nu\beta\beta$ from the 
pulse shape simulation~\cite{PSSpaper} and the one measured
with DEP events:~0.018.
\end{itemize}

The $0\nu\beta\beta$ survival fraction can be cross checked with the one
determined for $2\nu\beta\beta$ decays.  The energy region is chosen between
1. and 1.45~MeV to exclude the $\gamma$ lines at 1461~keV from $^{40}$K and
1525~keV from $^{42}$K.  The spectral decomposition of the BEGe
data~\cite{bckgpap} yields a fraction of $f_{2\nu\beta\beta} = 0.66\pm0.03$ of
$2\nu\beta\beta$ decays. The parts $f_i$ of the remaining components are
listed in Table~\ref{tab:2nbb} together with the PSD survival fractions
$\epsilon_i$.  The background origins mostly from Compton scattered $\gamma$
quanta.  The fractions $\epsilon_i$ were extrapolated from several studies
involving experimental measurements as well as simulations.  For $^{228}$Th,
$\epsilon_i$ is determined from present calibration data.

\begin{table}[b]
\begin{center}
   \caption{   \label{tab:2nbb}
        Decomposition of events in the region between 1~MeV and
        1.45~MeV. Listed are the estimated fraction $f_i$ \cite{bckgpap} and
        the total efficiency $\epsilon_i$ for each component $i$.  }
   \begin{tabular}[tbp]{lcc}\hline
   component & $f_i$ & $\epsilon_i$ \\  
\hline
   $^{40}$K                      & $0.032 \pm 0.009$ & $0.56 \pm 0.03$ \\  
   $^{42}$K in LAr               & $0.187 \pm 0.022$ & $0.49 \pm 0.05$ \\  
   $^{42}$K at {\it n+} surface  & $0.030 \pm 0.017$ & $0.30 \pm 0.04$ \\  
   $^{60}$Co                     & $0.013 \pm 0.013$ & $0.29 \pm 0.02$ \\  
   $^{60}$Co intrinsic           & $0.002 \pm 0.001$ & $0.21 \pm 0.02$ \\  
   $^{68}$Ga intrinsic           & $0.007 \pm 0.007$ & $0.33 \pm 0.02$ \\  
   $^{214}$Bi                    & $0.036 \pm 0.014$ & $0.41 \pm 0.02$ \\  
   $^{228}$Th                    & $0.003 \pm 0.002$ & $0.54 \pm 0.03$ \\  
   {\it p+} events              & $0.003 \pm 0.002$ & $0.02 \pm 0.02$ \\  
\hline
   other                        & $0.024 \pm 0.024$ & $0.45 \pm 0.45$\\ 
\hline
   \end{tabular}
\end{center}
\end{table}

The PSD survival fraction for $2\nu\beta\beta$ decays
$\epsilon_{2\nu\beta\beta}$ is then related to the overall PSD survival
fraction for events in the interval $\epsilon_{\rm data}$~=~0.748~$\pm$~0.011
(Table~\ref{tab:calib-cut}) by:
\begin{equation}
 \epsilon_{\rm data} = f_{2\nu\beta\beta}\cdot \epsilon_{2\nu\beta\beta} + 
                       \sum_i f_i \cdot \epsilon_i
\label{eq:2nubbeff}
\end{equation}

The resulting survival fraction of $2\nu\beta\beta$ events is
$\epsilon_{2\nu\beta\beta}$~=~0.90~$\pm$~0.05.  This number needs a small
correction due to decays in the {\it n+} transition layer.  The long pulse
rise time for these events (see Sect.~\ref{ssec:psdbege}) leads to a ballistic
deficit in the reconstructed energy, i.e.~$0\nu\beta\beta$ events do not
reconstruct at the peak position. This loss is already accounted for in the
definition of the dead layer thickness. For $2\nu\beta\beta$ events the energy
spectrum is continuous, i.e.~the effective dead volume is smaller. But $A/E$
is reduced as well and a fraction of about 0.015~$\pm$~0.005 is rejected
according to simulations. For the comparison with the $0\nu\beta\beta$ PSD
survival fraction, this correction should be added such that finally a
fraction of 0.91~$\pm$~0.05 is obtained. It agrees well with
$\epsilon_{0\nu\beta\beta}$~=~0.92~$\pm$~0.02.

\subsection{PSD summary for BEGe detectors}

Due to their small area {\it p+} contact BEGe detectors offer a powerful pulse
shape discrimination between $^{76}$Ge $0\nu\beta\beta$ signal events of
localized energy deposition and background events from multiple interactions
in the detector or energy deposition on the surface.

The parameter $A/E$ constitutes a simple discrimination variable with a clear
physical interpretation allowing a robust PSD analysis.  The characteristics
of this quantity have been studied for several years and are applied for the
first time in a $0\nu\beta\beta$ analysis. $^{228}$Th data taken once per week
are used to calibrate the performance of $A/E$ and to correct for the observed
time drifts and small energy dependencies.  The whole procedure of the PSD
analysis was verified using $2\nu\beta\beta$ events from $^{76}$Ge recorded
during physics data taking.

The chosen cut accepts a fraction of 0.92~$\pm$~0.02 of $0\nu\beta\beta$
events and rejects 33 out of 40 events in a 400~keV wide region around
$Q_{\beta\beta}$ (excluding the central 8~keV blinded window). The latter is
compatible with the expectation given our background composition and PSD
rejection.  The background index is reduced to
(0.007$_{-0.002}^{+0.004}$)~\ctsper.

Applying the PSD cut to $2\nu\beta\beta$ events results in an estimated
$0\nu\beta\beta$ signal survival fraction of 0.91~$\pm$~0.05 that agrees very
well with the value extracted from DEP and simulations.

\section{Pulse shape discrimination for semi-coaxial detectors}

In the current Phase~I analysis, three independent pulse shape selections have
been performed for the semi-coaxial detectors. They use very different
techniques but it turns out that they identify a very similar set of events as
background. The neural network analysis will be used for the $0\nu\beta\beta$
analysis while the other two (likelihood classification and PSD selection
based on the pulse asymmetry) serve as cross checks.

All methods optimize the event selection for every detector individually. They
divide the data into different periods according to the noise performance.
Two detectors (ANG~1 and RG~3) had high leakage current soon after the
deployment. The analyses discussed here consider therefore only the other six
coaxial detectors.

\subsection{Pulse shape selection with a neural network }

\begin{figure}[b]
   \begin{center}
      \includegraphics[width=0.95\columnwidth]{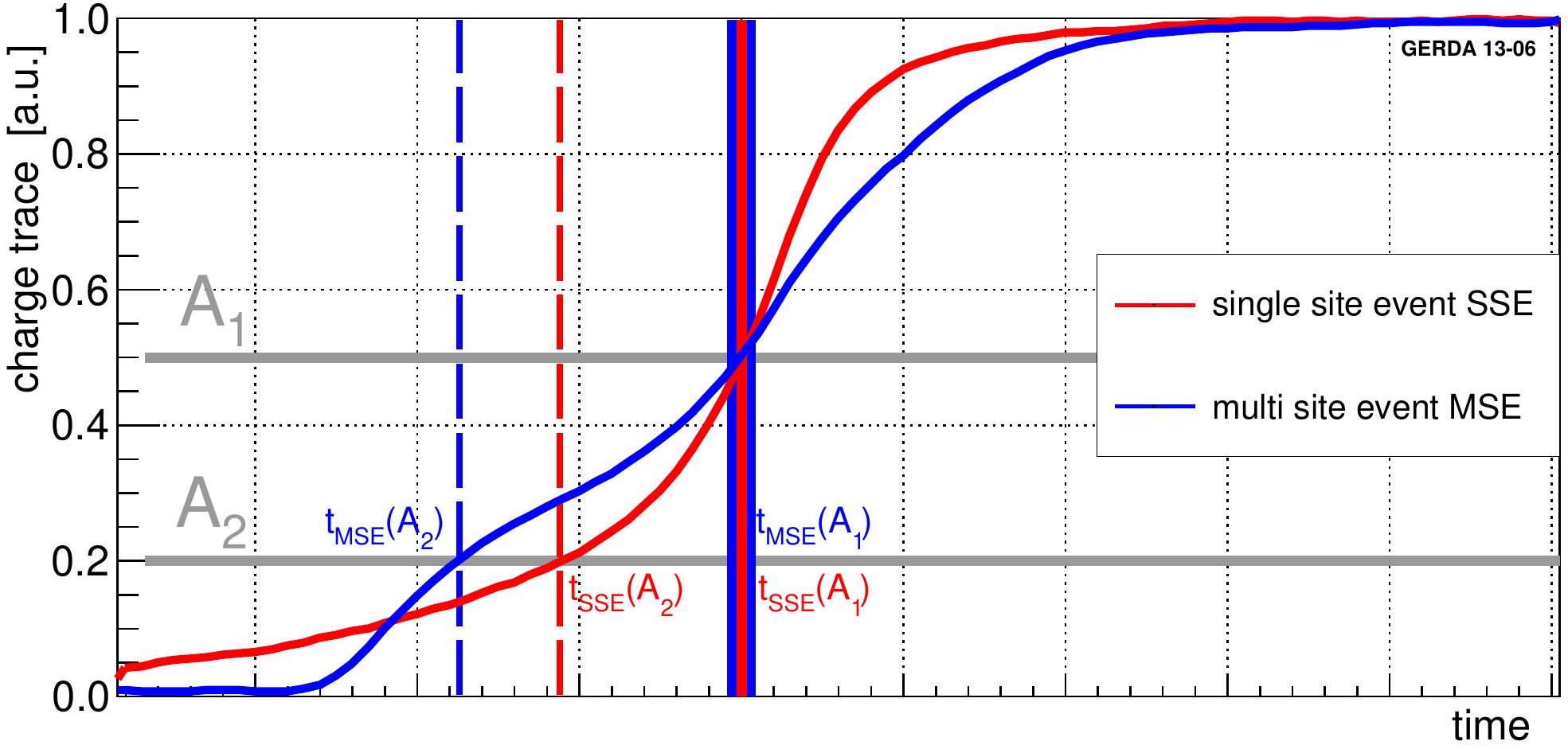}
   \caption{ \label{fig:amplitude}
                Example physics data pulses for SSE and MSE 
                candidate events.  The determination of the input
                parameters for the TMVA algorithms is shown for
                pulse heights $A_1$ and $A_2$.
   }
   \end{center}
\end{figure}

The entire current pulse or - to be more precise - the rising part of the
charge pulse is used in the neural network analysis.  The following steps are
performed to calculate the input parameters:
\begin{itemize}
\item baseline subtraction using the recorded pulse information in the
  80~\mus\ before the trigger. If there is a slope in the baseline due to
  pile up, the event is rejected. This selection effects practically only
  calibration data,
\item smoothing of the pulse with a moving window averaging of 80~ns
  integration time,
\item normalization of the maximum pulse height to one to remove the energy
  dependence,
\item determination of the times when the pulse reaches 1, 3, 5, ..., 99\,\%
  of the full height. The time when the pulse height reaches $A_1$= 50\,\%
  serves as reference.  Due to the 100~MHz sampling frequency, a (linear)
  interpolation is required between two time bins to determine the
  corresponding time points (see Fig.~\ref{fig:amplitude}).
\end{itemize}

The resulting 50 timing informations of each charge pulse are used as input to
an artificial neutral network analyses.  The TMVA toolkit implemented in
ROOT~\cite{tmva} offers an interface for easy processing and evaluation.  The
selected algorithm TMlpANN~\cite{TMlpANN} is based on multilayer
perceptions. Two hidden layers with 51 and 50 neurons are used.  The method is
based on the so called ``supervised learning'' algorithm.
 
\begin{figure}[t]
   \begin{center}
      \includegraphics[width=0.95\columnwidth]{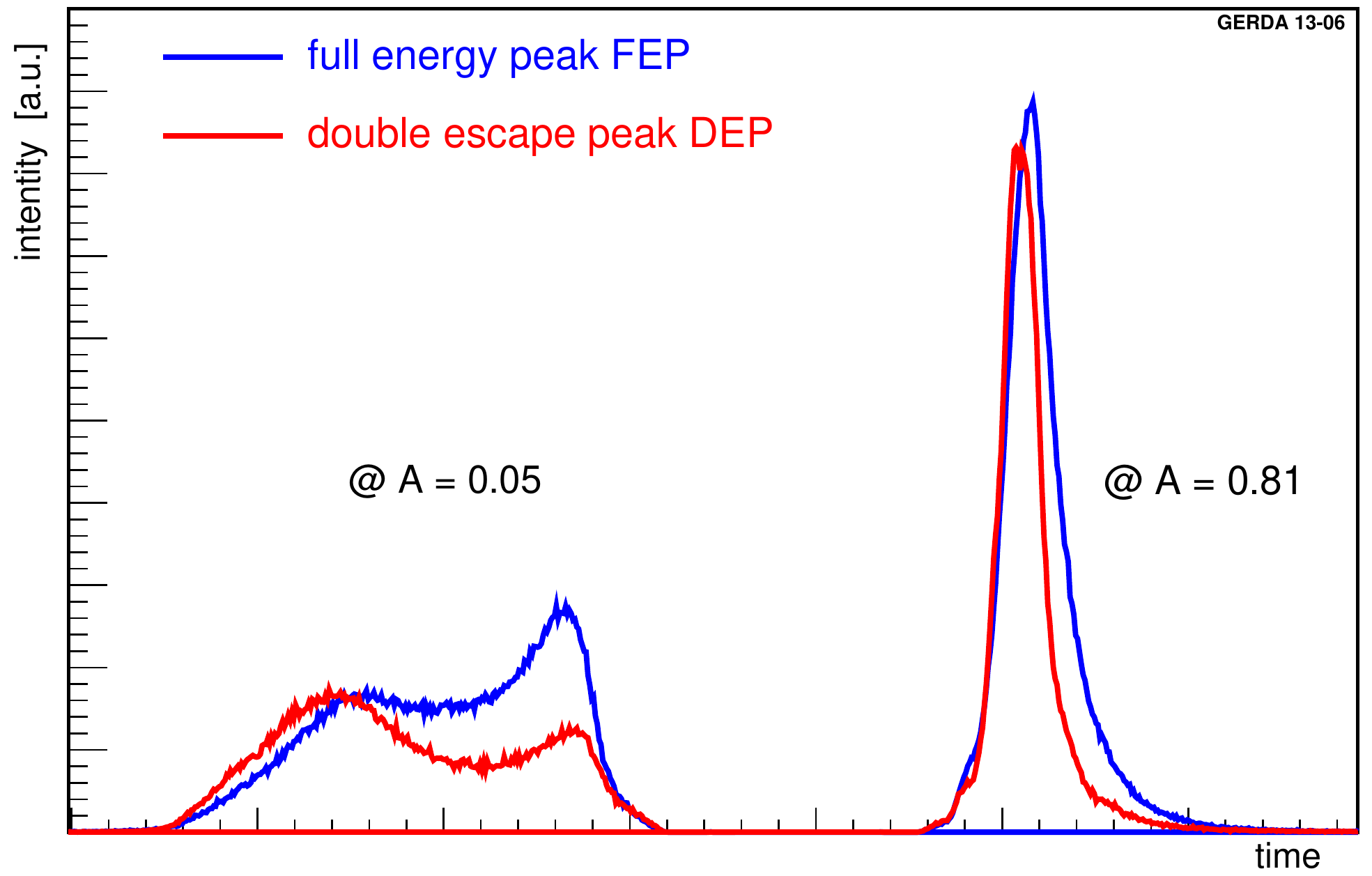}
   \caption{ \label{fig:variables_5_81}
     Time distribution for crossing the 5\,\% (left) and 81\,\% (right)
     pulse height for $^{228}$Th calibration events with energy close
     to the DEP (red) and close to the 1621~keV FEP (blue).
   }
   \end{center}
\end{figure}

Calibration data are used for training.  DEP events in the interval $1593$~keV
$\pm1\cdot$FWHM serve as proxy for SSE while events of the full energy line of
$^{212}$Bi in the equivalent interval around $1621$~keV are dominantly MSE and
are taken as background sample.  Fig.~\ref{fig:variables_5_81} shows as an
example of the separation power the distribution of the time of 5\,\% and
81\,\% pulse height for the two event classes. Note that both event classes
are not pure samples but a mixture of SSE and MSE because of the Compton
events under the peaks.

The calibrations are grouped in three intervals.  The first period spans from
the start of data taking to July 2012 when the detector configuration and some
electronics was changed (p1).  The second period (p2) lasts the first four
weeks afterwards and the third period (p3) the rest of Phase~I. For RG~2, the
second period spans until November 2012 when its operating voltage was
reduced.  For each period at least 5000 events are available per detector and
event class for training.

\begin{figure}[t]
   \begin{center}
      \includegraphics[width=0.95\columnwidth]{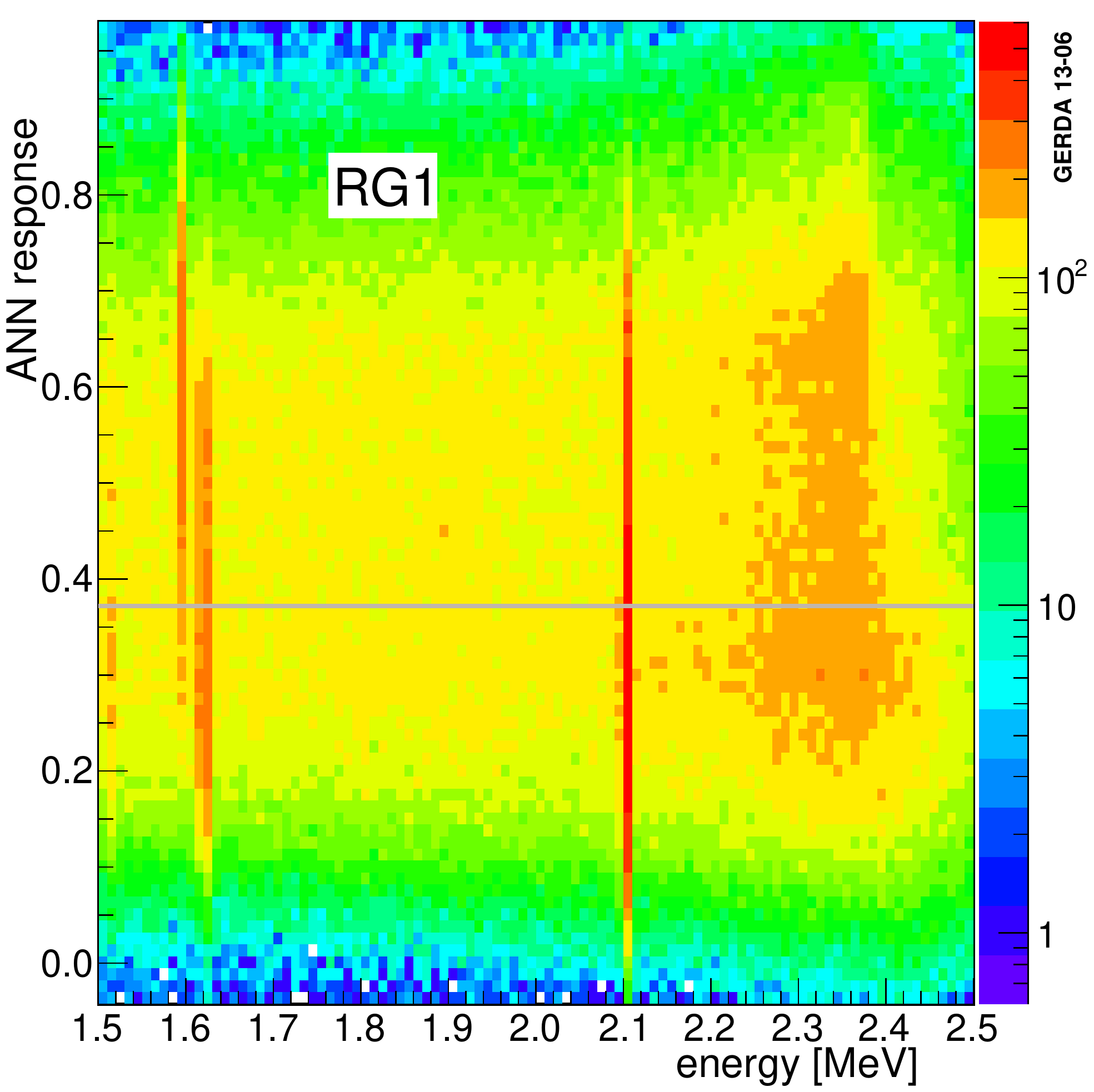}
   \caption{ \label{fig:tmlp_en}
          TMlpANN response versus energy for $^{228}$Th calibration
          events. Shown is the distribution for RG~1. The line
         at $\sim$0.38 marks the position for 90\,\% DEP survival
         fraction.
 }
   \end{center}
\end{figure}

\begin{figure}[b]
   \begin{center}
      \includegraphics[width=0.95\columnwidth]{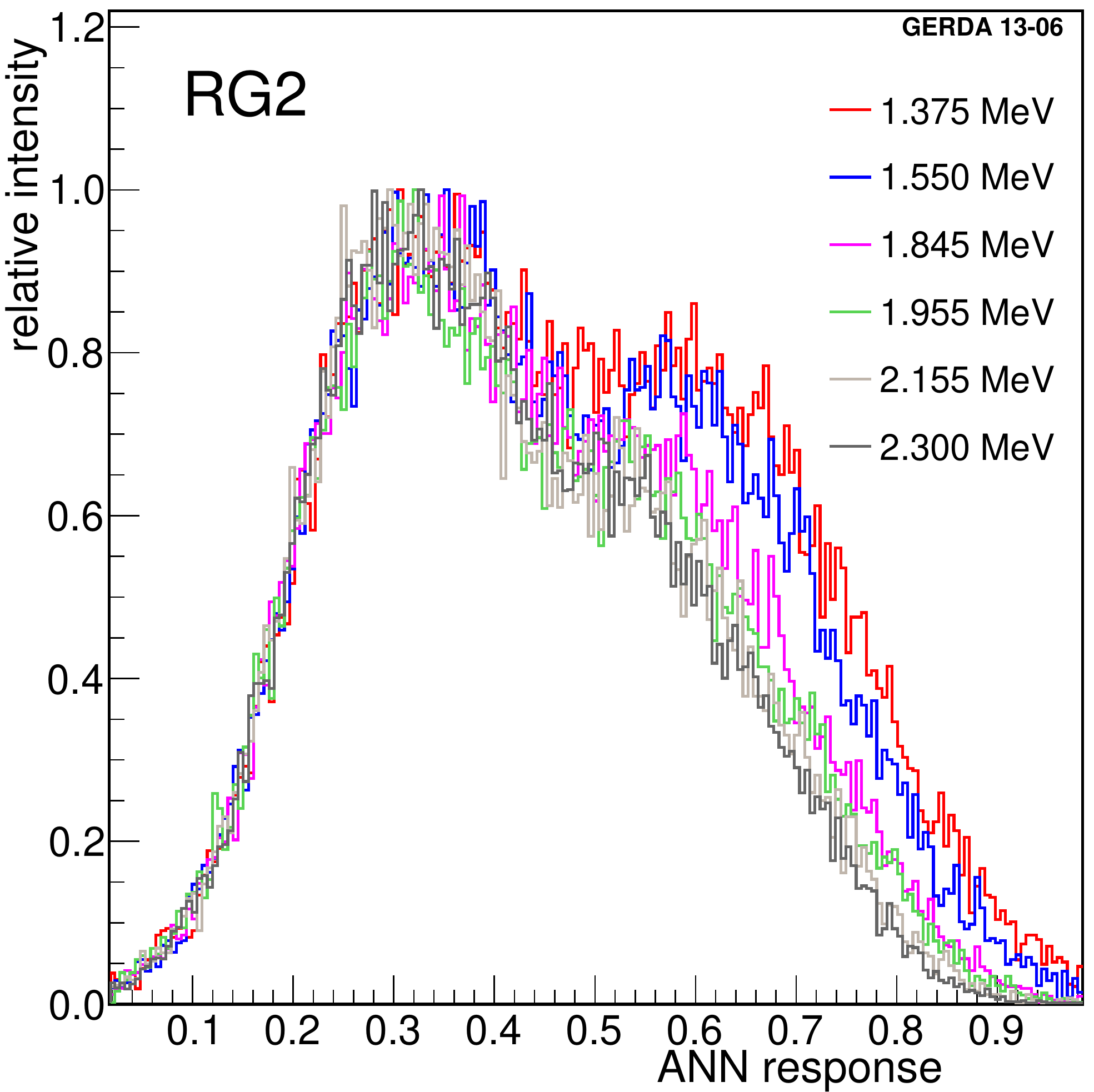}
   \caption{ \label{fig:comptondrift}
          TMlpANN response for Compton events for RG~2 at different energies.
          The energy dependence for RG~2 is about twice bigger than for any
          other detector.
   }
   \end{center}
\end{figure}

The output of the neural network is a qualifier, i.e.~a number between
$\approx 0$ (background like event) and $\approx 1$ (signal like event).
Fig.~\ref{fig:tmlp_en} shows a scatter plot of this variable versus the
energy. The distribution peaks for DEP events at higher qualifier values while
for FEP events at 1621~keV and SEP events at 2104~keV the intensity is shifted
to lower values.  The qualifier distribution from Compton events at different
energies can be compared to estimate a possible energy dependence of the
selection (see Fig.~\ref{fig:comptondrift}). For most detectors no drift is
visible. Only RG~2 shows a larger variation. An energy dependent empirical
correction of the qualifier is deduced from such distributions.

\begin{figure}[t]
   \begin{center}
      \includegraphics[width=0.9\columnwidth]{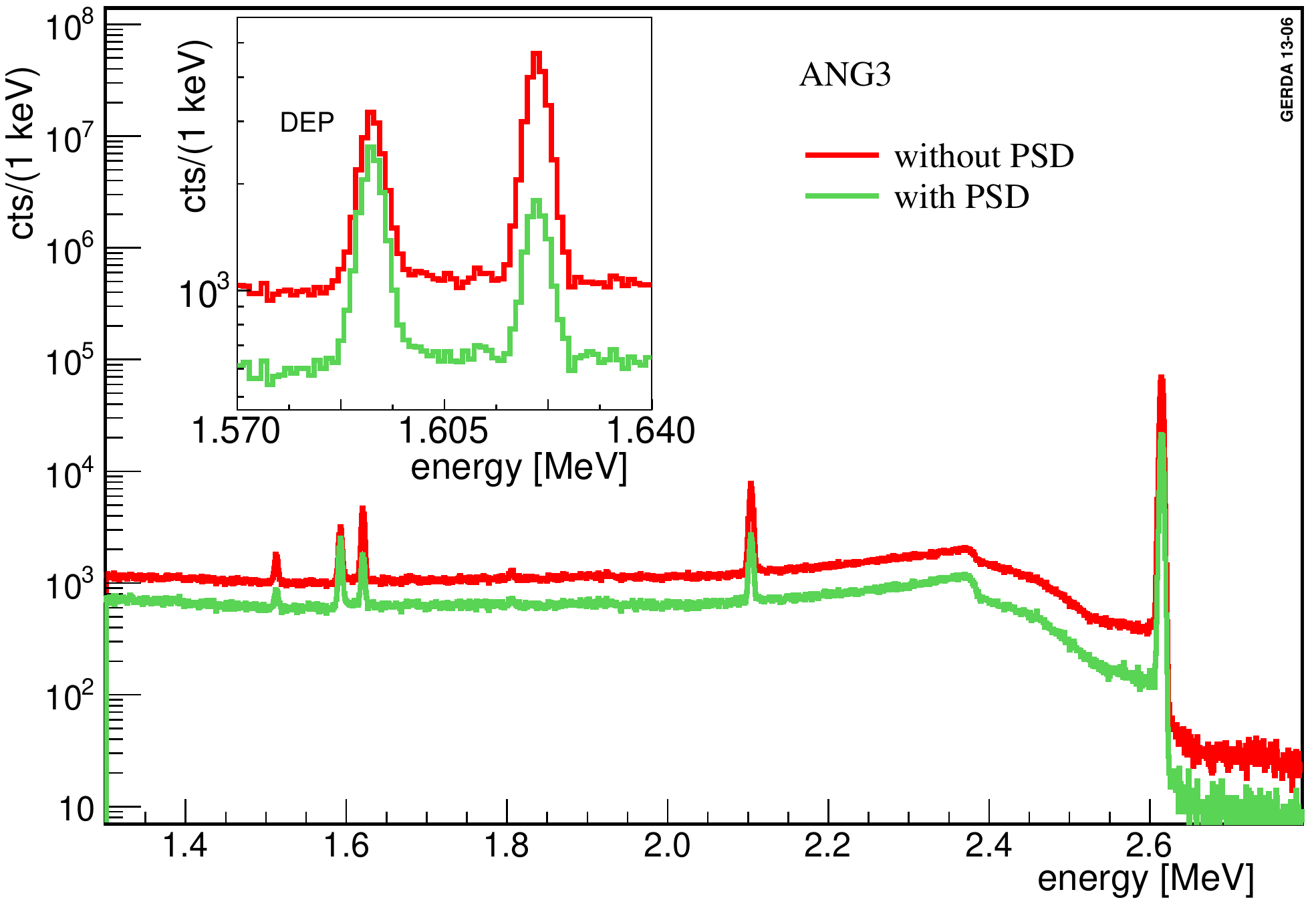}
   \caption{ \label{fig:coax_th_spectrum}
          $^{228}$Th calibration spectrum without and with TMlpANN pulse shape
          discrimination for ANG~3. The PSD cut is fixed to retain  90\,\% of
          DEP events (see inset). 
}
   \end{center}
\end{figure}

The qualifier threshold which keeps 90\,\% of the DEP events is determined for
each detector and each period individually.  The cut values vary between 0.31
and 0.42.  Fig.~\ref{fig:coax_th_spectrum} shows a $^{228}$Th calibration
spectrum with and without PSD selection.  For the analysis, the survival
fraction of MSE is studied. The survival is defined as the fraction of the
peak content remaining after the cut, i.e.~the Compton events under the peak
are subtracted by scaling linearly the event counts from energies below and
above the peak.  The fractions are listed in Table~\ref{tab:coax_psd} for the
different periods.  The last column lists the number of events in the 230~keV
window around $Q_{\beta\beta}$ before and after the cut. About 45\,\% of the
events are classified as background.

\begin{table}[b]
\begin{center}
\caption{\label{tab:coax_psd}
           Survival fractions of the neural network PSD for different event
           classes and different detectors.  Numbers are given for calibration
           (cal.) or physics data from the periods p1, p2 and
           p3. The statistics of physics data for p2 are
           small and hence not always listed. ``$2\nu\beta\beta$'' stands for
           the 1.0 - 1.4~MeV interval which consists dominantly of
           $2\nu\beta\beta$ decays. $^{42}$K signifies the 1525~keV full
           energy peak.  ROI is here the 230 keV window around
           $Q_{\beta\beta}$. The errors are typically 0.01 for SEP and ROI for
           calibration, 0.02 for the $2\nu\beta\beta$ data interval and 0.06
           for the $^{42}$K $\gamma$ peak.  The last column list the event
           count after/before the PSD cut.
}
\begin{tabular}{lcccccc}
det. & period &  SEP & ROI & $2\nu\beta\beta$ & $^{42}$K & ROI\\ 
\hline & & cal. & cal. & data & data & data\\ \hline 
ANG~2 & p1  & 0.33 & 0.58 & 0.74 & 0.30 & 2/4\\ 
ANG~2 & p2  & 0.50 & 0.65 & 0.65 & & 0/1\\ 
ANG~2 & p3  & 0.47 & 0.63 & 0.73 & 0.40 & 6/8\\ 
ANG~3 & p1  & 0.32 & 0.56 & 0.79 & 0.43 & 6/9\\ 
ANG~3 & p2  & 0.34 & 0.56 & 0.75 & & 2/3\\ 
ANG~3 & p3  & 0.40 & 0.63 & 0.82 & 0.44 & 4/6\\ 
ANG~4 & p1  & 0.29 & 0.54 & 0.78 & 0.45 & 1/1\\ 
ANG~4 & p2  & 0.28 & 0.53 & 0.63 & & 0/1\\ 
ANG~4 & p3  & 0.33 & 0.58 & 0.83 & 0.44 & 2/4\\ 
ANG~5 & p1  & 0.26 & 0.55 & 0.79 & 0.41 & 2/11\\ 
ANG~5 & p2  & 0.21 & 0.45 & 0.57 & & 0/2\\ 
ANG~5 & p3  & 0.33 & 0.59 & 0.80 & 0.30 & 6/16\\ 
RG~1 & p1  & 0.45 & 0.63 & 0.80 & 0.52 & 2/6\\ 
RG~1 & p2  & 0.43 & 0.60 & 0.77 & & 2/3\\ 
RG~1 & p3  & 0.41 & 0.62 & 0.81 & 0.48 & 3/4\\ 
RG~2 & p1  & 0.30 & 0.53 & 0.82 & 0.49 & 10/12\\ 
RG~2 & p2  & 0.37 & 0.60 & 0.81 & 0.48 & 3/3\\ 
RG~2 & p3  & 0.45 & 0.61 & 0.76 & 0.56 & 2/2\\
\end{tabular}
\end{center}
\end{table}

\begin{figure}[htbp]
   \begin{center}
      \includegraphics[width=0.95\columnwidth]{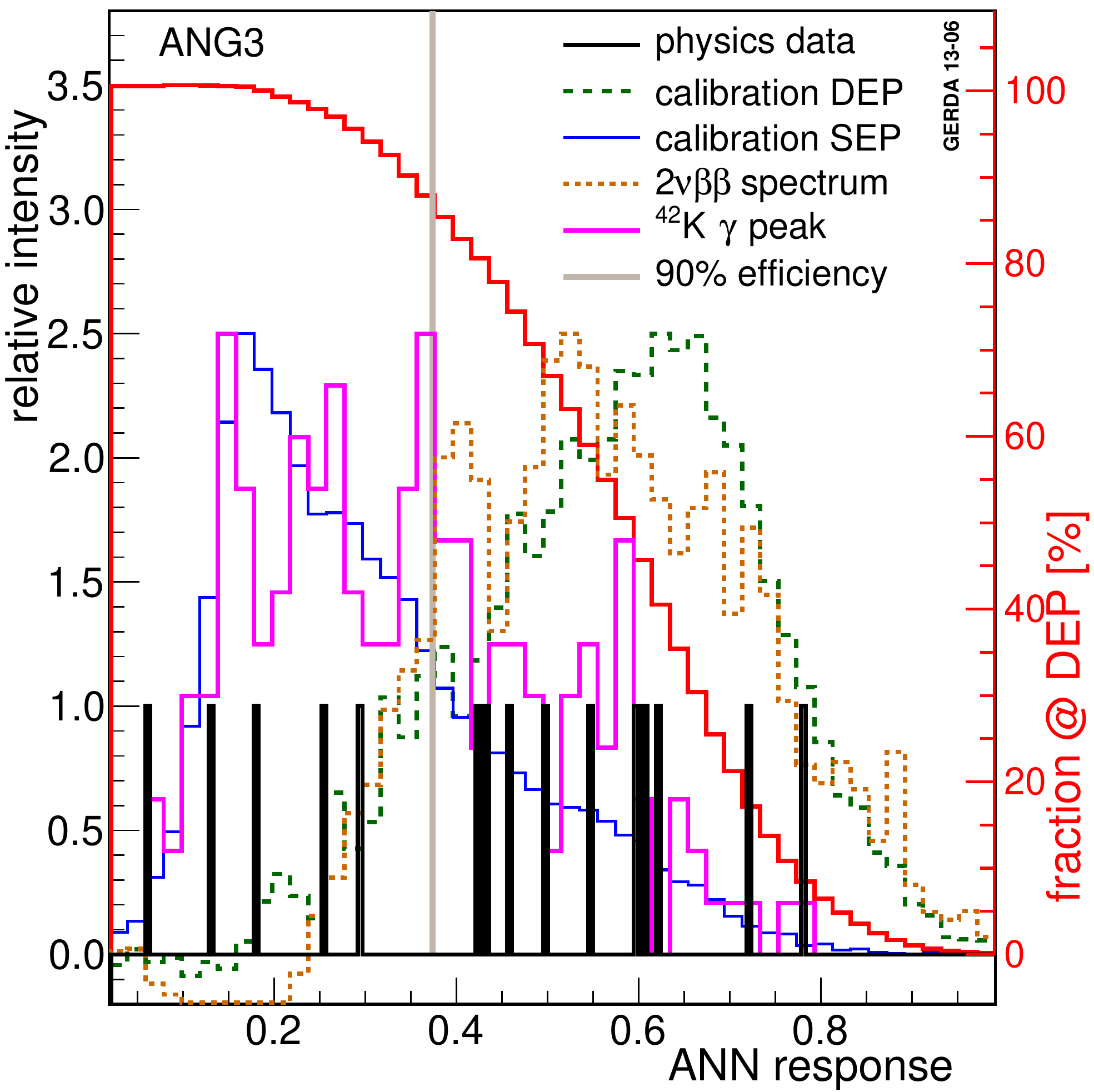}
   \caption{ \label{fig:qualifier_1}
         ANN response for $^{228}$Th calibration events for DEP (green, long
         dashes) and SEP (dark blue) for ANG~3 in the first period.  The
         distributions from Compton events at these energies are subtracted
         statistically using events in energy side bands.  Also shown in black
         are the qualifier values of events from physics data taking from a
         230~keV window around $Q_{\beta\beta}$. The grey vertical line marks
         the cut position.  Physics data events from the 1525~keV FEP of
         $^{42}$K are shown in magenta and the ones from the interval 1.0 -
         1.4~MeV by brown dashes (dominantly $2\nu\beta\beta$, MSE part
         subtracted).
}
   \end{center}
\end{figure}

Fig.~\ref{fig:qualifier_1} shows the ANN response for DEP and SEP events.
Shown are also the qualifier distributions for different samples from physics
data taking: from the interval 1.0 - 1.4~MeV (dominantly $2\nu\beta\beta$
events, MSE part subtracted), from the 1525~keV $^{42}$K $\gamma$ line
(dominantly MSE) and the qualifier for events in the 230~keV window.  The
events from the 1525~keV gamma peak are predominantly MSE and the shape agrees
with the SEP distribution. The events in the 1.0 - 1.4 MeV region are
dominantly SSE and their distribution agrees quite well with the one for DEP
events.  The red curve shows the DEP survival fraction versus the cut position
(right scale).

\begin{figure*}[tbp]
   \begin{center}
      \includegraphics[width=0.9\textwidth]{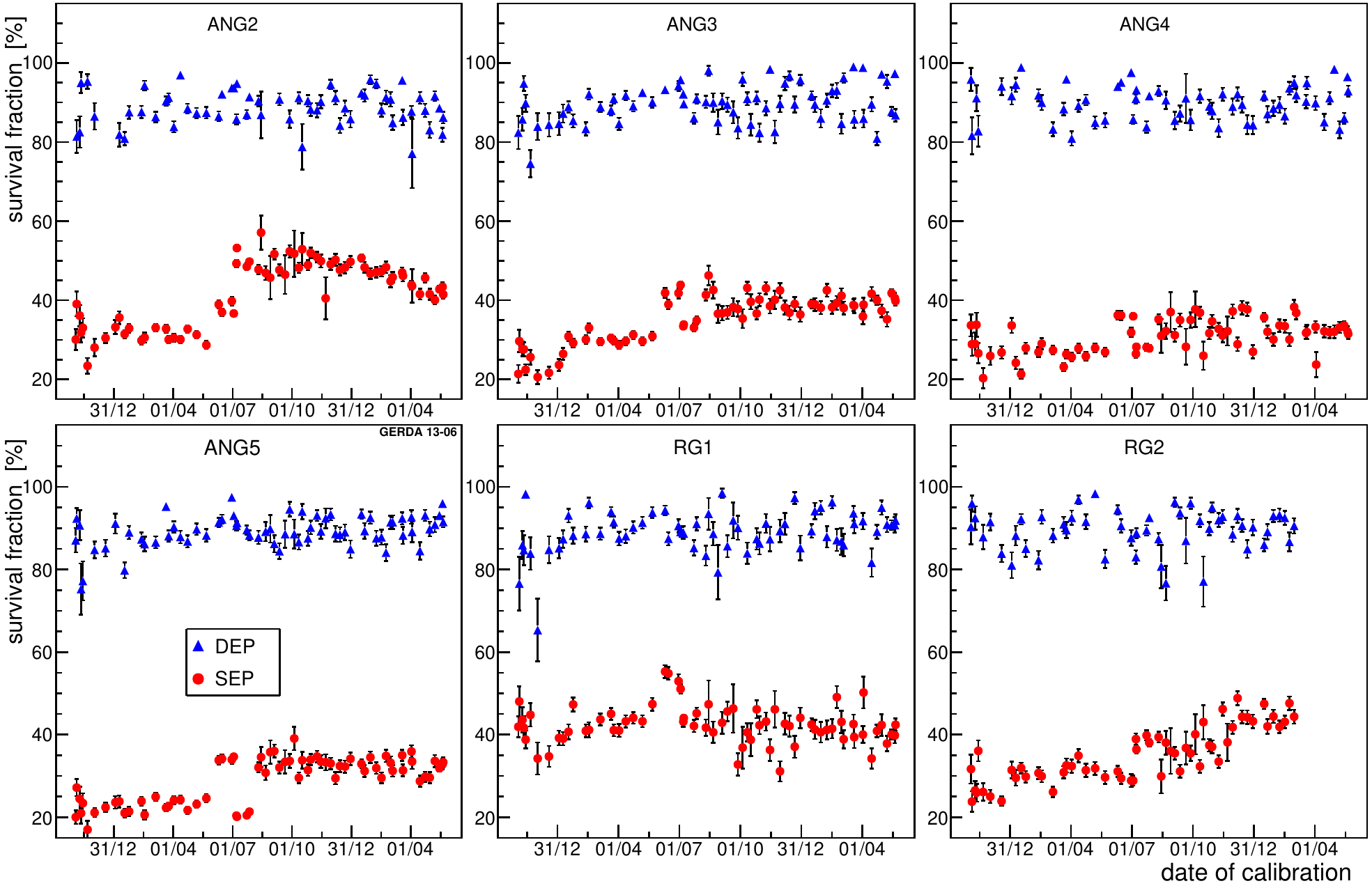}
   \caption{ \label{fig:nn_time_SEP}
          DEP (blue) and SEP (red) survival fraction  for individual
          calibrations for the entire Phase~I.
  }
   \end{center}
\end{figure*}

The training was performed for the periods individually by combining all
calibration data.  The rules can then be applied to every single calibration
to look for drifts in time. Fig.~\ref{fig:nn_time_SEP} shows the DEP survival
fraction (blue triangles) for the entire Phase~I from November 2011 to May
2013 for all detectors. The plots show a stable performance.  Also shown are
the equivalent entries (red circles) for events with energy around the SEP
position.  For several detectors the rejection of MSE is not stable.
Especially visible is the deterioration starting in July 2012.  This is
related to different conditions of high frequency noise.

\begin{figure}[btp]
   \begin{center}
      \includegraphics[width=0.9\columnwidth]{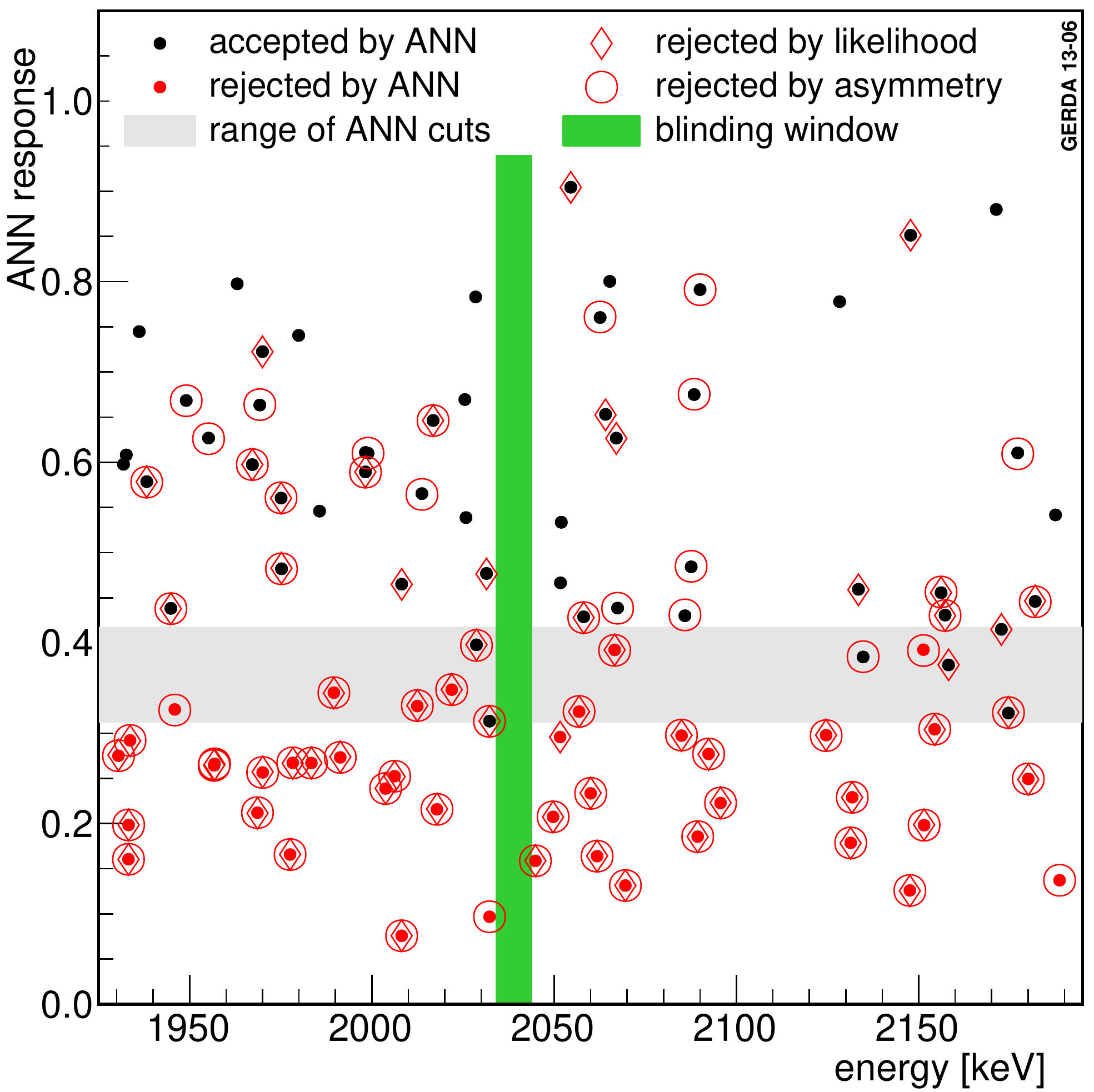}
   \caption{ \label{fig:ann_roi}
        Neural network qualifier for  events with energy close to \qbb. Events
        marked by a red dot are rejected. Circles and diamonds mark events
        which are rejected by the likelihood analysis and the method based on
        the  pulse asymmetry, respectively.
 }
   \end{center}
\end{figure}

The distribution of the qualifier for all events in the 230~keV window around
$Q_{\beta\beta}$ is shown in Fig.~\ref{fig:ann_roi}.
Events rejected by the neural network are marked in red.  Circles mark events
rejected by the likelihood method and diamonds those rejected by the method
based on the current pulse asymmetry.  Both methods are discussed below. In
the shown energy interval, all events removed by the neural network are also
removed by at least one other method and for about 90\,\% of the cases, all
three methods discard the events. In a larger energy range about 3\,\% of the
rejected events are only identified by the neural network.
\begin{figure}[btp]
   \begin{center}
      \includegraphics[width=0.9\columnwidth]{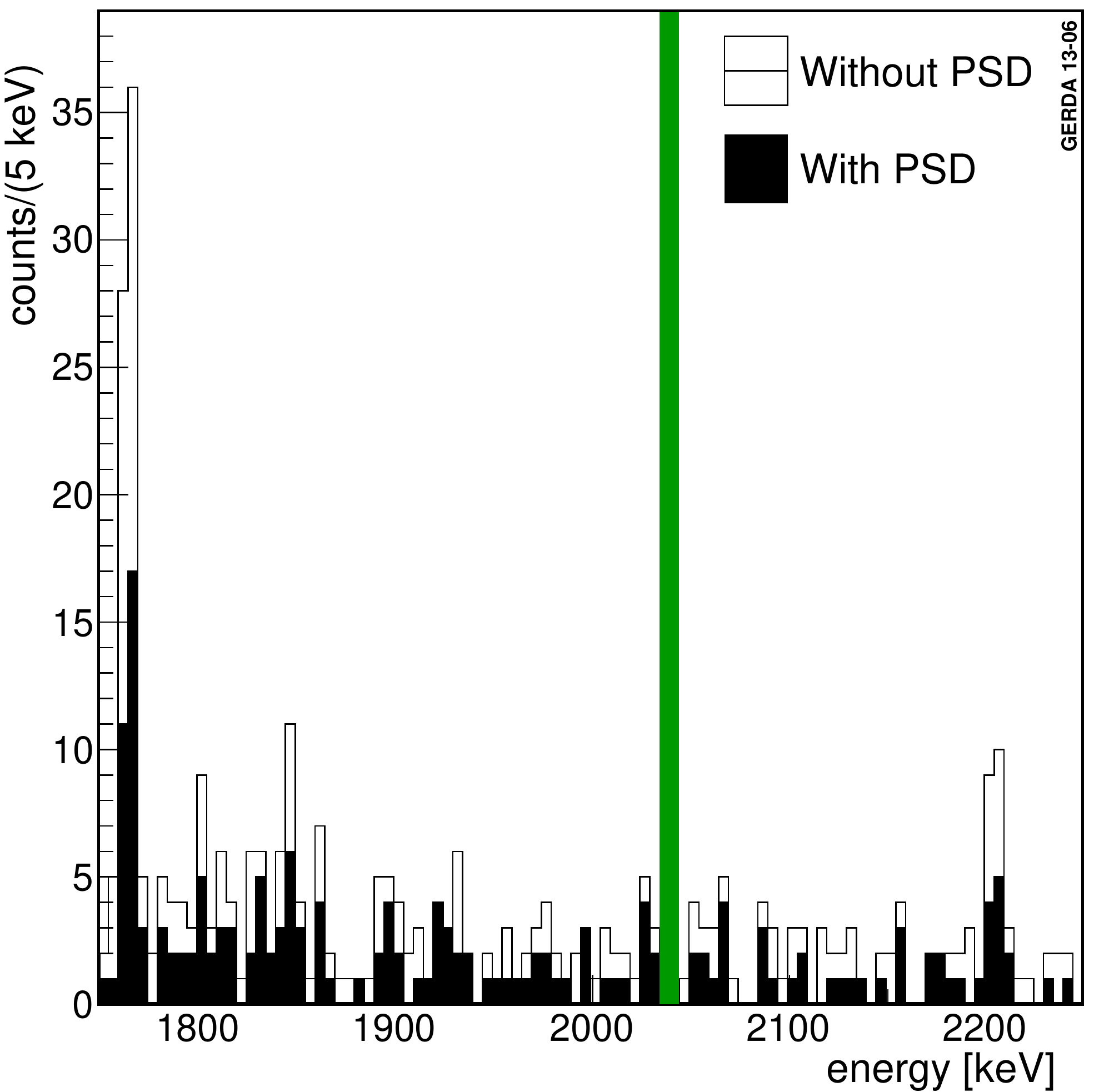}
   \caption{ \label{fig:eneplo}
            Energy spectrum of semi-coaxial detectors with and without
            neural network PSD selection.
 }
   \end{center}
\end{figure}

Fig.~\ref{fig:eneplo} shows the energy spectrum of all semi-coaxial detectors
added up before and after the PSD selection.

\subsection{Systematic uncertainty of the neural network signal efficiency}

In this analysis we use the survival fraction of DEP events as efficiency for
$0\nu\beta\beta$ events.

The distribution of DEP events in a detector is not homogeneous since the
probability for the two 511~keV photons to escape is larger in the corners. It
is therefore conceivable that the ANN - instead of selecting SSE - is mainly
finding events at the outer surface.  The DEP survival fraction would in this
case not represent the efficiency for $0\nu\beta\beta$ decay which are
distributed homogeneously in the detector.

$2\nu\beta\beta$ events are also SSE and homogeneously distributed inside the
detector.  Hence a comparison of its pulse shape identification efficiency
with the preset 0.90 value for DEP events is a powerful test.

Another SSE rich sample are events at the Compton edge of the 2614.5~keV
$\gamma$ line. The energy range considered is 2.3 - 2.4~MeV, i.e.~higher than
$Q_{\beta\beta}$. The comparison to the DEP survival fraction allows also to
check for an energy dependence.  The distribution of Compton edge events in
detector volume is similar to DEP.

\subsubsection{Efficiency of $2\nu\beta\beta$ for neural network PSD}

The energy range between 1.0 and 1.3~MeV (position of the Compton edge of the
1525~keV line) is suited for the comparison of the SSE efficiency. At lower
energies the electronic noise will deteriorate the discrimination between SSE
and MSE. In this interval, the data set consists to a fraction
$f_{2\nu\beta\beta}=0.76\pm0.01$ of $2\nu\beta\beta$ decays according to the
\GERDA\ background model~\cite{bckgpap}.  The remaining 24\,\% are Compton
events predominantly of the 1525~keV line from $^{42}$K decays, of the
1460~keV line from $^{40}$K decays and from $^{214}$Bi decays.  Hence it is a
good approximation to use the pulse shape survival fraction $\epsilon_{\rm
  Compton}$ from the calibration data to estimate the suppression of the
events not coming from $2\nu\beta\beta$ decays. Typical values for
$\epsilon_{\rm Compton}$ are between 0.6 and 0.7 for the different detectors,
i.e.~higher than the values quoted in Table~\ref{tab:coax_psd} due to a small
energy dependence (see Fig.~\ref{fig:coax_th_spectrum}).

\begin{figure}[t]
   \begin{center}
      \includegraphics[width=0.95\columnwidth]{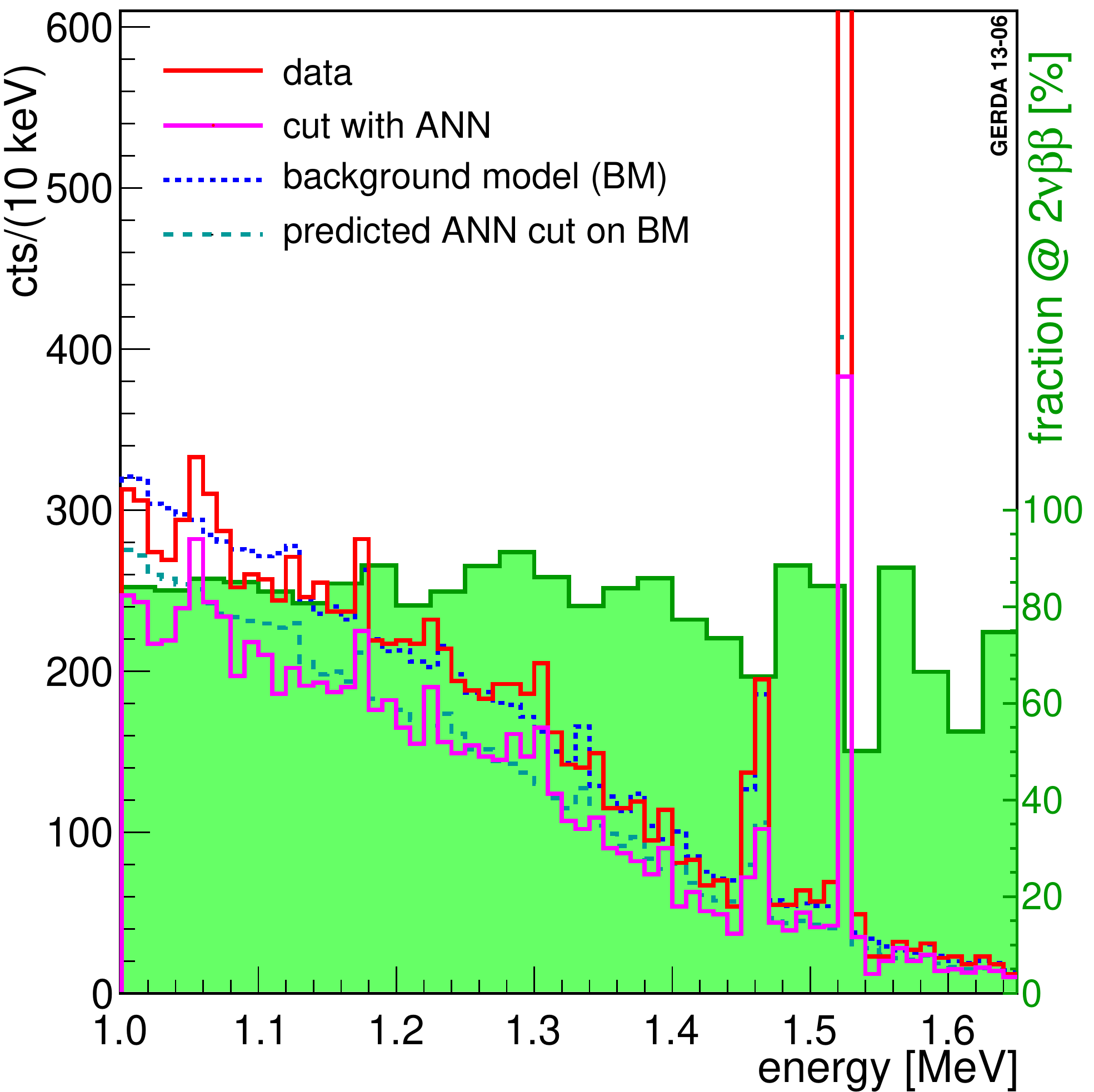}
   \caption{ \label{fig:eff_2nubb}
           Effect of the PSD selection on the data (in red and magenta) and
           the expected effect on the background model (dark blue dotted and
           light blue dashed). Overlayed is also the extracted PSD efficiency
           (green filled histogram) for $2\nu\beta\beta$ events (right side
           scale).
}
   \end{center}
\end{figure}

Fig.~\ref{fig:eff_2nubb} shows the physics data (red) overlayed with the
background model (blue, taken from Ref.~\cite{bckgpap}) and the same
distributions after the PSD cut (in magenta for the data and in light blue for
the model). For the model, the $2\nu\beta\beta$ fraction is scaled by the DEP
survival rate while the remaining fraction is scaled according to
$\epsilon_{\rm Compton}$ taken from the $^{228}$Th calibration data for each
detector. Both pairs of histograms agree roughly in the range 1.0 - 1.3~MeV.
This is qualitatively confirmed if the $2\nu\beta\beta$ PSD efficiency is
calculated using \eqref{eq:2nubbeff}.  Its distribution is also shown as the
green filled histogram in Fig.~\ref{fig:eff_2nubb}. The average efficiency for
the range 1.0 - 1.3~MeV is $\epsilon_{2\nu\beta\beta} = 0.85\pm 0.02$ where
the error is dominated by the systematic uncertainty of $\epsilon_{\rm
  Compton}$.  The latter is estimated by a variation of the central value by
10\,\% which is the typical variation of $\epsilon_{\rm Compton}$ between
1~MeV and 2~MeV.

The obtained efficiency $\epsilon_{2\nu\beta\beta}$ is close to the DEP
survival fraction of $\epsilon_{\rm DEP} =0.9$ and indicates that there are no
sizable systematic effects related to the differences in the distribution of
DEP and $2\nu\beta\beta$ events in the detectors.

\subsubsection{Neural  network PSD survival fraction of Compton edge events}

Calibration events at the Compton edge of the 2615~keV $\gamma$ line, i.e.~in
the region close to 2.38~MeV, are enhanced in SSE and distributed similar to
DEP events in the detector. The qualifier distribution for these events can be
approximated as a linear combination of the DEP distribution and the one from
multiple Compton scattered $\gamma$ ray events (MCS).  Events with energy
larger than the Compton edge (e.g. in the interval 2420 - 2460~keV) consists
almost exclusively of MCS.  The total counts in the qualifier interval 0 to
0.2 for Compton edge events and MCS are used for normalization and the MCS
distribution is then subtracted.

\begin{figure}[bp]
   \begin{center}
      \includegraphics[width=0.95\columnwidth]{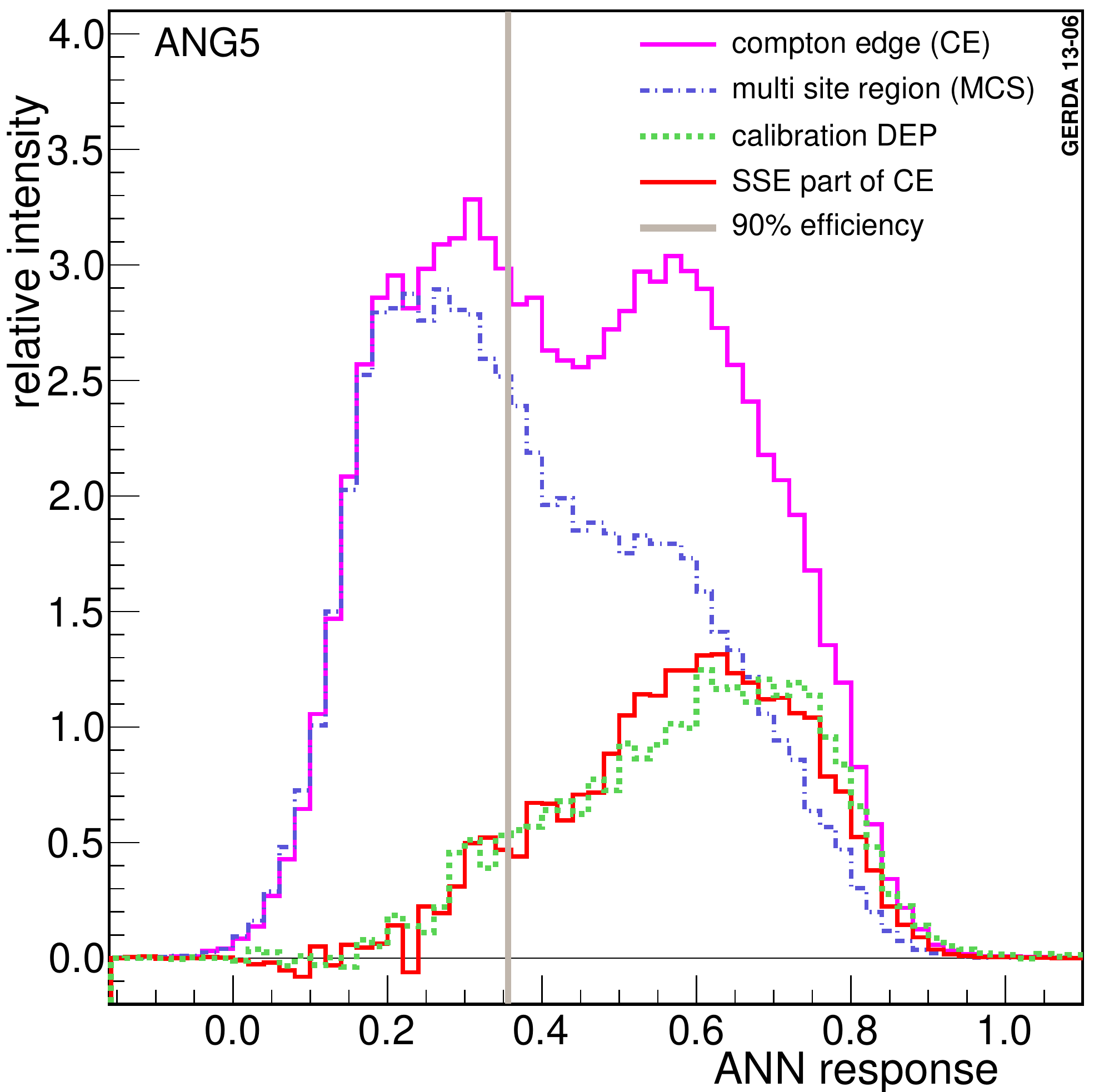}
   \caption{ \label{fig:comptonedge}
          Qualifier distribution for events at the Compton edge (magenta) as a
          linear combination of MCS (blue) and DEP (green dotted)
          distributions.  The Compton edge distribution after the subtraction
          of the SEP part is shown in red.
}
   \end{center}
\end{figure}

The ``MCS subtracted'' Compton edge distribution (red curve in
Fig.~\ref{fig:comptonedge}) shows an acceptable agreement with the DEP
distribution (green dotted curve).  The survival fraction is defined as the
part above the selection cut.  Its value varies for the 3 periods and the 6
detectors between 0.85 and 0.94. No systematic shift relative to the DEP value
e.g.~due to an energy dependence of the efficiency is visible. If SEP events
are used to model the multi site event contribution, consistent values are
obtained.

\subsubsection{Summary of systematic uncertainties}

The cross checks of the PSD efficiency address a possible energy dependence
and a volume effect due to the different distributions of DEP and
$0\nu\beta\beta$ events. All studies performed are based on calibration or
physics data and are hence independent of simulations.

The possible deviations from 0.90 seen are combined quadratically and scaled
up to allow for additional sources of systematic uncertainties.  The
$0\nu\beta\beta$ efficiency is $\epsilon_{\rm ANN}=0.90^{+0.05}_{-0.09}$.


\subsection{Alternative PSD methods}

Two more PSD methods have been developed.  They are used here to cross check
the event selection of the neural network method (see Fig.~\ref{fig:ann_roi}).
No systematic errors for the signal efficiency has been evaluated for them.

\subsubsection{Likelihood analysis }

In a second PSD analysis, 8 input variables calculated from the charge pulse
trace are used as input to the projective likelihood method implemented in
TMVA.  Each input variable is the sum of four consecutive pulse heights of
10~ns spacing after baseline subtraction and normalization by the energy. The
considered trace is centered around the time position where the derivative of
the original trace is maximal, i.e.~around the maximum of the current.

The training is performed for two periods: before (pI) and after (pII) June
2012. Instead of DEP events, the Compton edge in the interval 2350 - 2370~keV
is used as signal region and the interval 2450 - 2570~keV as background
sample. The latter contains only multiple Compton scattered photons and is
hence almost pure MSE. The Compton edge events are a mixture of SSE and MSE.
From the two samples a likelihood function for signal $L_{\rm sig}$ and
background $L_{\rm bkg}$ like events is calculated and the qualifier
$q_{\rm PL}$ is the ratio $ q_{\rm PL} = L_{\rm sig}/(L_{\rm sig} + L_{\rm bkg})$.

\begin{figure}[b]
   \begin{center}
      \includegraphics[width=0.95\columnwidth]{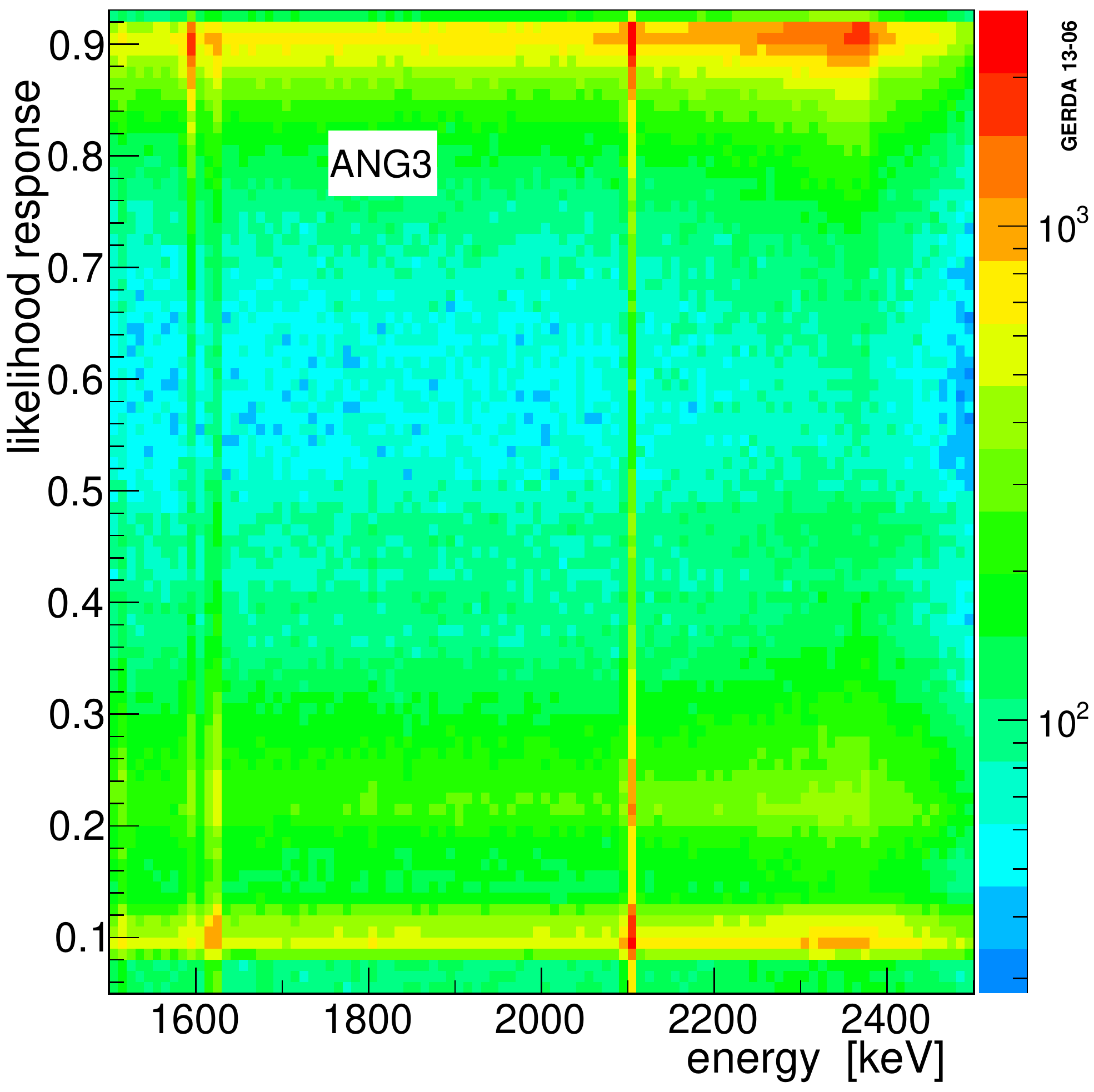}
   \caption{         \label{fig:pl_qualifier_e}
               Likelihood response versus energy distribution for $^{228}$Th
               calibration events. Data are shown for ANG~3.
}
   \end{center}
\end{figure}

\begin{table}[t]
\begin{center}
\caption{\label{tab:pl_eff} 
     Survival fractions of the projective likelihood PSD for different event
     classes and the different detectors. The cut for each subset is set to
     yield a DEP survival fraction of 0.8.  Numbers are given for calibration
     data (cal.) or physics data. pI and pII indicate the two
     periods. The meaning of the columns are identical to
     Table~\ref{tab:coax_psd} and the same applies to the size of statistical 
     errors for the different samples. 
}
\begin{tabular}{lcccccc} \hline
det. & period & SEP & ROI & $2\nu\beta\beta$ & $^{42}$K & ROI\\ 
     &     & cal. & cal.  & data & data & data \\ \hline 
ANG~2 & pI  & 0.47 & 0.57 & 0.61 & 0.35 &1/3 \\ 
ANG~2 & pII & 0.50 & 0.56 & 0.57 & 0.37 & 4/10 \\ 
ANG~3 & pI  & 0.49 & 0.58 & 0.60 & 0.36 & 2/7\\ 
ANG~3 & pII & 0.52 & 0.61 & 0.64 & 0.40 & 3/11 \\ 
ANG~4 & pI  & 0.52 & 0.60 & 0.65 & 0.54 & 1/1 \\ 
ANG~4 & pII & 0.50 & 0.62 & 0.71 & 0.51 & 2/5 \\ 
ANG~5 & pI  & 0.45 & 0.57 & 0.62 & 0.42 & 0/8 \\ 
ANG~5 & pII & 0.40 & 0.51 & 0.61 & 0.31 & 3/21\\ 
RG~1 & pI   & 0.50 & 0.63 & 0.63 & 0.59 & 2/6 \\ 
RG~1 & pII  & 0.51 & 0.62 & 0.65 & 0.46 & 2/7 \\ 
RG~2 & pI   & 0.49 & 0.60 & 0.70 & 0.46 & 6/8\\ 
RG~2 & pII  & 0.51 & 0.61 & 0.63 & 0.50 & 7/9 \\ \hline
\end{tabular}
\end{center}
\end{table}

Fig.~\ref{fig:pl_qualifier_e} shows for the calibration data the scatter plot
of the qualifier versus energy. The separation of DEP (1593~keV) and FEP at
1621 keV is visible by the different population densities at low and high
qualifier values.  The cut position is independent of energy and fixed to
about $0.80$ survival fraction for DEP events.  The SEP survival fractions and
for comparison also the ones for several other subsets are listed in
Table~\ref{tab:pl_eff}. About 65\,\% of the events in the 230~keV window
around $Q_{\beta\beta}$ are rejected.

\begin{figure}[t]
   \begin{center}
      \includegraphics[width=0.95\columnwidth]{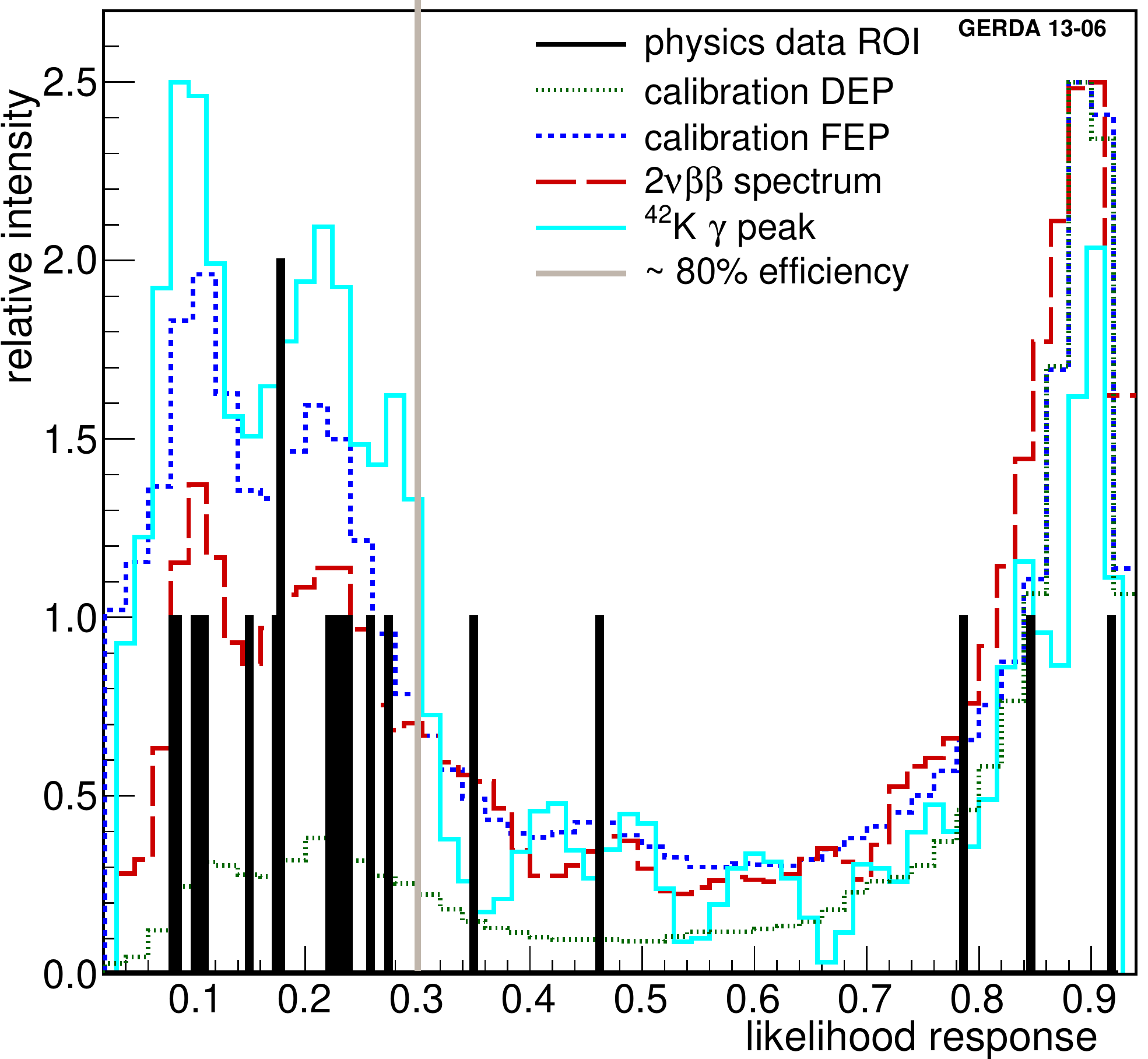}
   \caption{ \label{fig:pl_qualifier}
           Likelihood response for $^{228}$Th calibration DEP (green dotted)
           and FEP (dark blue dashed) events for ANG~3. The distributions from
           Compton events at these energies are subtracted statistically using
           events in energy side bands.  Also shown in black are the qualifier
           values of events from physics data taking from a 230~keV window
           around $Q_{\beta\beta}$. The grey vertical line marks the cut
           position.  Shown are also distributions of physics data events from
           the $^{42}$K $\gamma$ line (light blue) and from the interval 1.0 -
           1.4~MeV (red, dominantly $2\nu\beta\beta$).
}
   \end{center}
\end{figure}

Fig.~\ref{fig:pl_qualifier} shows the distribution of the qualifier for
different event classes. The distribution for physics data events from the
$^{42}$K line are well described by the FEP distribution in calibration data
and the events in the 1.0 - 1.4~MeV interval are clearly enhanced in SSE as
expected for $2\nu\beta\beta$ events.

\subsubsection{PSD based on pulse asymmetry }

In a third approach, only two variables are used to select single site events
for the semi-coaxial detectors.  As discussed above, the $A/E$ variable alone
is not a good parameter for semi-coaxial detectors. However, if $A/E$ is
combined with the pulse asymmetry, the PSD selection is much more
effective. The asymmetry $A_s$ is defined as
\begin{equation}
  A_s = \frac{ \Sigma_{i=0}^{i=n_m} I(i) - \Sigma_{i=n_m}^{i<200} I(i)
  } { \Sigma_{i=0}^{i<200} I(i) }
\end{equation}
Here $I(i)$ is the current pulse height, i.e.~the differentiated charge pulse
at time $i$, and $n_m$ the time position of the maximum. A window of 200
samples (i.e.~a 2~\mus\ time interval) around the time of the trigger is
analyzed.

To reduce noise, different moving window averaging with integration times of 0
(no filter), 20, 40, 80, 160 and 320~ns for the charge pulse are applied. For
each shaping time, $A/E$ and $A_s$ are determined. Empirically, the
combination
\begin{equation}
q_{AS}=A/E \cdot (c+A_s)
\end{equation}
exhibits good PSD performance.  For SSE, the current pulse might contain more
than one maximum (Fig.~\ref{fig:coax_sse}).  To reduce ambiguities, $A_S$ is
shaped with larger integration times.

\begin{figure}[t]
   \begin{center}
      \includegraphics[width=0.95\columnwidth]{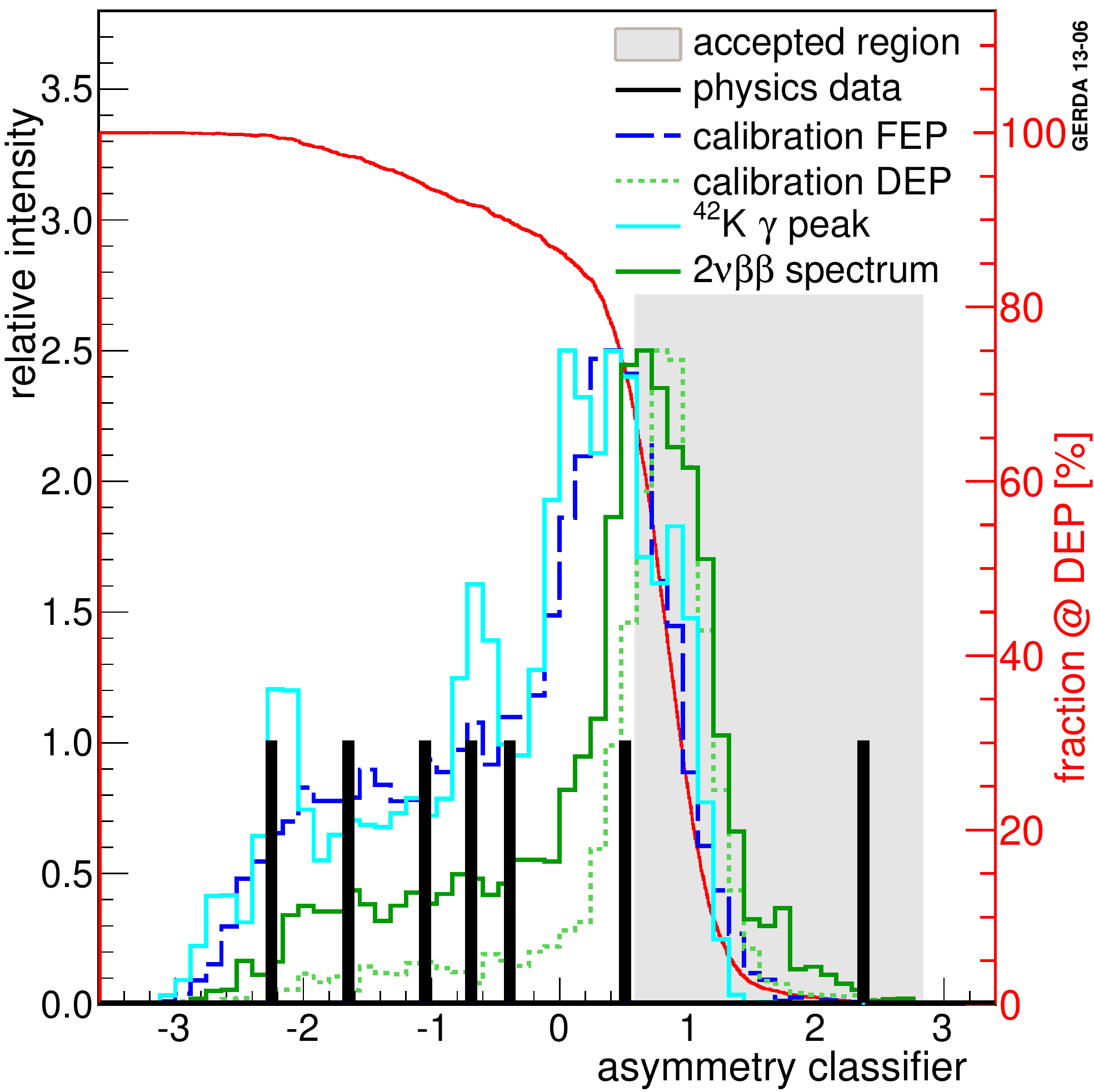}
   \caption{ \label{fig:thomas_1d}
           Distribution of qualifier for DEP (dotted green) and 
           FEP (dashed dark blue)
           calibration events for ANG~3 after a statistical subtraction of the
           Compton events below the peaks. The grey band marks the acceptance
           range.  Overlayed are also the PSD qualifier for physics data in the
           230~keV window around $Q_{\beta\beta}$ (black), data events from
           the 1525~keV $^{42}$K peak (light blue) and from the interval 1.0 -
           1.4 MeV (dark green dotted).  The DEP survival fraction is
           displayed in red (right scale).
}
   \end{center}
\end{figure}

An optimization is performed by comparing the DEP survival fraction
$\epsilon_{DEP}$ from calibration data to the fraction of background events
$f_{\rm bkg}$ between 1700 and 2200~keV (without a 40~keV blinded interval
around $Q_{\beta\beta}$) that remains after the PSD selection.  The lower cut
value of the qualifier $q_{AS}$ is determined by maximizing the quantity
$S=\epsilon_{DEP} / \sqrt{f_{\rm bkg} + 3/N_{\rm bkg}}$; the upper cut is
fixed at $\approx +4\sigma$ of the Gaussian width of the DEP qualifier
distribution (see Fig.~\ref{fig:thomas_1d}). All combinations of shaping times
for $A/E$ and $A_s$ are scanned as well as different values for $c$ in the
range of 1 - 4. The one with the highest $S$ is selected.

The term $3/N_{\rm bkg}$ with $N_{\rm bkg}$ being the total number of
background events is added to avoid an optimization for zero background.  For
$N_{\rm bkg} \approx 40$ the optimization yields a DEP survival fraction of
0.7 - 0.9 (see Table~\ref{tab:coax_psd2}) and about 75\,\% of the events in
the interval 1.7 - 2.2~MeV are rejected.

\begin{figure}[b]
   \begin{center}
      \includegraphics[width=0.95\columnwidth]{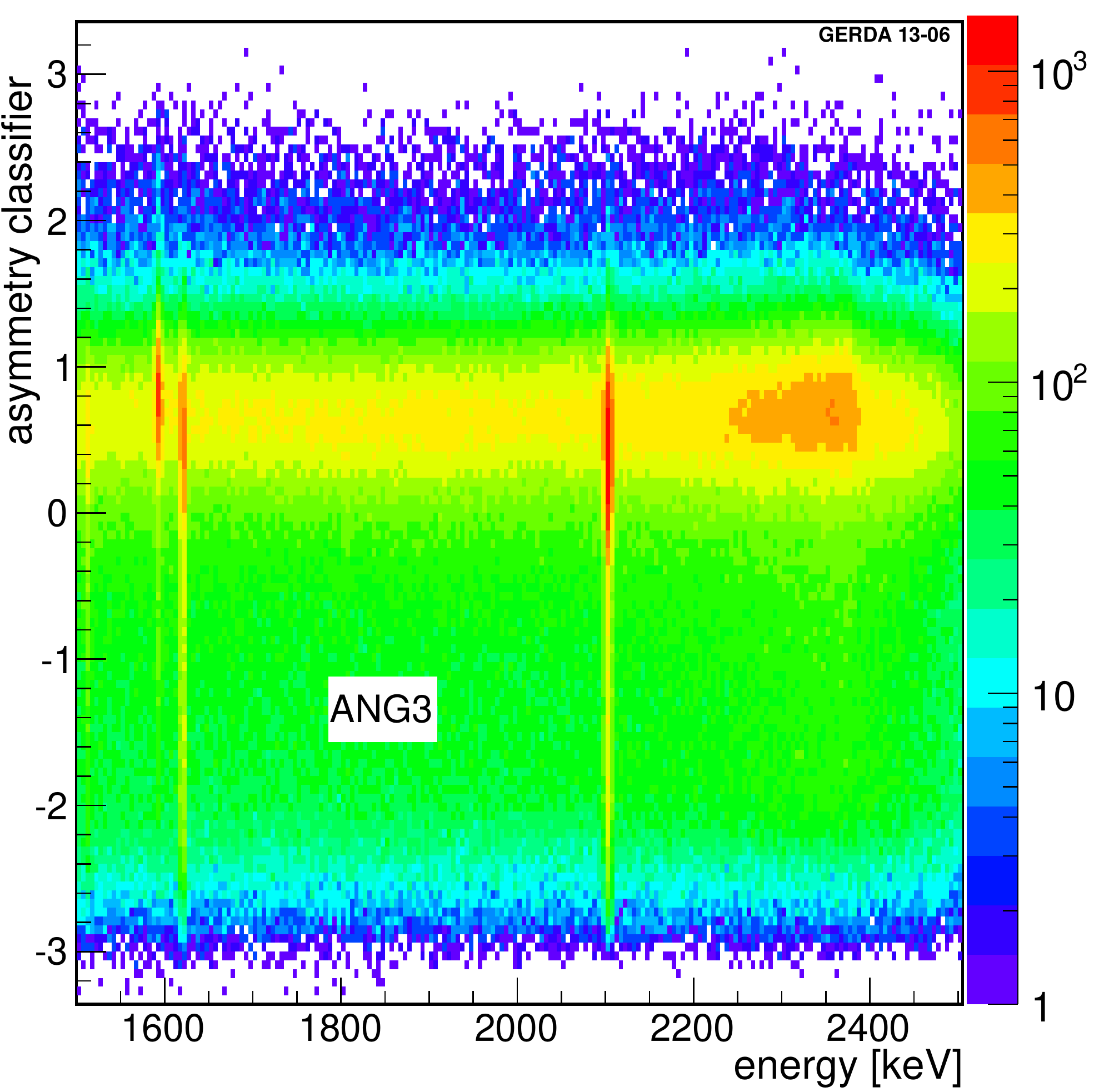}
   \caption{ \label{fig:thomas_2d}
           Distribution of the ANG~3 qualifier versus energy for $^{228}$Th
           calibration data for the PSD based on the pulse asymmetry.
     }
   \end{center}
\end{figure}

Fig.~\ref{fig:thomas_2d} shows a scatter plot of the PSD qualifier versus the
energy. A separation between the DEP and multi site events at the energy of
the FEP or SEP is visible.
Fig.~\ref{fig:thomas_1d} shows qualifier distributions for DEP and FEP
calibration events after Compton events below the peaks are statistically
subtracted.  Overlayed is also the PSD qualifier for physics data in the
230~keV window around $Q_{\beta\beta}$ (black histogram), from the 1525~keV
$\gamma$ line (light blue) and the interval 1.0 - 1.4~MeV (yellow).  The right
scale shows the DEP survival fraction (red) as a function of the cut
position. The grey area indicates the accepted range.  The qualifier
distribution of physics data around $Q_{\beta\beta}$ has a larger spread than
the one of FEP events. This is the reason why events at $Q_{\beta\beta}$ are
rejected stronger than MSE (see Table~\ref{tab:coax_psd2}).  A possible
explanation is that the physics data contain a large fraction of events which
are not MSE. These can be for example surface {\em p+} events.  The
``maximal'' background model of \GERDA~\cite{bckgpap} is compatible with a
significant fraction of {\em p+} events.  A pulse shape simulation also shows
that the selection corresponds to a volume cut: events close to the {\it p+}
contact and in the center of the detectors are removed.
\begin{table}[bth]
\begin{center}
\caption{\label{tab:coax_psd2}
    Survival fractions of the PSD based on the current pulse asymmetry for
    different event classes and the different detectors.  Numbers are given
    for calibration data (cal.) or physics data. pI and pII stand for the two
    periods.  The DEP survival fractions are listed in the third column.  Note
    that the selection of data files is slightly different for this analysis
    such that the total observed event counts (last column) are different
    compared to the other PSD methods. The meaning of the different columns is
    explained in Table~\ref{tab:coax_psd} and the same applies to the size of
    statistical errors for the different samples.
}
\begin{tabular}{lcccccc}
det. & time  & DEP & SEP & $2\nu\beta\beta$ & $^{42}$K & ROI \\
 & &  cal. & cal.  & data & data & data \\ \hline 
ANG~2 & pI  & 0.69 & 0.32  & 0.52 & 0.28 & 1/5 \\
ANG~2 & pII & 0.70 & 0.40  & 0.50 & 0.33 & 4/6 \\
ANG~3 & pI  & 0.90 & 0.51  & 0.74 & 0.55 & 3/13 \\
ANG~3 & pII & 0.69 & 0.22  & 0.49 & 0.23 & 1/7 \\
ANG~4 & pI  & 0.78 & 0.28  & 0.63 & 0.41 & 1/9 \\
ANG~4 & pII & 0.78 & 0.45  & 0.66 & 0.41 & 2/8 \\
ANG~5 & pI  & 0.81 & 0.33  & 0.65 & 0.39 & 2/13 \\
ANG~5 & pII & 0.67 & 0.16  & 0.65 & 0.39 & 2/8 \\
RG~1 & pI   & 0.92 & 0.64  & 0.78 & 0.65 & 2/9 \\
RG~1 & pII  & 0.69 & 0.23  & 0.55 & 0.38 & 3/6 \\
RG~2 & pI   & 0.86 & 0.38  & 0.71 & 0.44 & 2/11 \\
RG~2 & pII  & 0.86 & 0.38  & 0.65 & 0.56 & 1/6 \\
\end{tabular}
\end{center}
\end{table}

\subsection{Summary of PSD analysis for coaxial detectors}

For the semi-coaxial detectors three different PSD methods are presented
following quite different concepts. The one based on an artificial neural
network will be used for the $0\nu\beta\beta$ analysis.  It has been tuned to
yield 90\,\% survival fraction for DEP events of the 2.6~MeV $\gamma$ line of
$^{208}$Tl decays. Most of these events are SSE like $0\nu\beta\beta$ decays.
For the study of a possible volume effect and energy dependence of the
efficiency, $2\nu\beta\beta$ decays ($\epsilon_{2\nu\beta\beta}=0.85\pm0.02$)
and events with energy close the Compton edge (efficiency between 0.85 and
0.95) have been used. We conclude that the $0\nu\beta\beta$ efficiency is
$\epsilon_{\rm ANN} = 0.90^{+0.05}_{-0.09}$.

The event selection of the neural network is cross checked by two other
methods.  One is based on a likelihood ratio. Training is performed with
events at the Compton edge (SSE rich) and at slightly higher energies (almost
pure MSE).  For a cut with a DEP survival fraction of about 0.8 only 45\,\% of
the events around $Q_{\beta\beta}$ remain.

Another method is only based on the $A/E$ parameter and the current pulse
asymmetry $A_S$. Different signal shapings are tried and an optimization of a
signal over background ratio is performed. The DEP survival fraction varies
between 0.7 and 0.9 for the different detectors and periods. The background is
reduced by a factor of four.

Of the events rejected by the neural network analysis in the 230~keV window
around $Q_{\beta\beta}$, about 90\,\% are also identified as background by
both other methods.  This gives confidence that the classification is
meaningful.

\section{Summary}

The neural network analysis rejects about 45\,\% of the events around
$Q_{\beta\beta}$ for the semi-coaxial detectors and the $A/E$ selection
reduces the corresponding number for BEGe detectors by about 80\,\%. With a
small loss in efficiency the \gerda\ background index is hence reduced from
$(0.021\pm0.002)$~\ctsper\ to $(0.010\pm0.001)$~\ctsper.  These values are the
averages over all data except for the period p2, the ``silver'' data set, that
covers the time period around the BEGe deployment and which corresponds to
6\,\% of the Phase~I exposure~\cite{bckgpap}.

The estimated $0\nu\beta\beta$ decay signal efficiencies for semi-coaxial
detectors are $0.90^{+0.05}_{-0.09}$ and for BEGe detectors $0.92\pm0.02$.
Despite this loss of efficiency, the \GERDA\ sensitivity defined as the
expected median half life limit of the $0\nu\beta\beta$ decay improves by
about 10\,\% with the application of the pulse shape discrimination.

\section*{Acknowledgments}
 The \gerda\ experiment is supported financially by
   the German Federal Ministry for Education and Research (BMBF),
   the German Research Foundation (DFG) via the Excellence Cluster Universe,
   the Italian Istituto Nazionale di Fisica Nucleare (INFN),
   the Max Planck Society (MPG),
   the Polish National Science Centre (NCN),
   the Foundation for Polish Science (MPD programme),
   the Russian Foundation for Basic Research (RFBR), and
   the Swiss National Science Foundation (SNF).
 The institutions acknowledge also internal financial support.

The \gerda\ collaboration thanks the directors and the staff of the LNGS
for their continuous strong support of the \gerda\ experiment.

We acknowledge guidance concerning the {\it n+} surface layer 
modeling from D.~Radford.

\end{document}